\documentclass[aps,prc,twocolumn,superscriptaddress]{revtex4}

\usepackage{amsmath}  
\usepackage{amsfonts}
\usepackage{amssymb}
\usepackage{natbib} 
\usepackage{setspace}
\usepackage{graphicx} 
\usepackage{xspace}
\usepackage{cancel}
\usepackage{url}
\usepackage{color}
\usepackage{float}

\newcommand{\eqn}[1]{Eq.~(\ref{#1})}
\newcommand{\fig}[1]{Fig.~\ref{#1}}
\newcommand{\tab}[1]{Table~\ref{#1}}

\newcommand{\sect}[1]{Section~\ref{#1}}

\usepackage{ulem}
\usepackage{color}
\def\bm{\boldmath}

\begin{document}

\title{
\vspace{-50mm}
  \begin{flushright}
    LFTC-23-6/79
  \end{flushright}
  \vspace{10mm}
$\eta$ and $\eta'$ mesons in nuclear matter and nuclei}

\author{J.~J.~Cobos-Mart\'inez\footnote{Corresponding Author. email:
\tt{jesus.cobos@unison.mx}}}
\affiliation{Departamento de F\'isica, Universidad de Sonora, Boulevard
Luis Encinas J. y Rosales, Colonia Centro, Hermosillo, Sonora 83000, M\'exico}
\author{Kazuo Tsushima}
\affiliation{Laboratório de Física Teórica e Computacional-LFTC, 
Universidade Cidade de S\~ao Paulo, 01506-000 S\~ao Paulo, SP, Brazil}

\date{\today}

\begin{abstract}
We present updated and extended results for the $\eta$-- and $\eta'$--nucleus bound
state energies, obtained by solving the Schr\"{o}dinger and Klein-Gordon equations with
 complex optical potentials, for a wide range of nuclei.
The $\eta$ and $\eta'$ nuclear potentials are obtained in the local density approximation
from  the mass shift  of these mesons in nuclear matter, which is calculated within the 
quark-meson coupling model.
Our results show that the $\eta$ and $\eta'$ mesons are expected to form mesic nuclei with all
the nuclei considered.
However, the signal for the formation of the $\eta$- and $\eta'$ mesic nuclei may be difficult
to identify experimentally due to possible large widths.
\end{abstract}


\maketitle

\date{\today}

\section{Introduction}

The investigation of how the properties and structure of hadrons, such as masses and widths,
are modified in a nuclear medium is one of the most exciting and important
problems in hadron and nuclear physics, since these are connected to, for example, chiral
 symmetry restoration and the structure of the QCD vacuum~\cite{piAF:2022gvw}, which in
  turn are reflected on the properties of hadrons in 
  medium~\cite{Metag:2017yuh,Krein:2017usp}.

In particular, research on the $\eta$ and $\eta'$ meson masses and widths in nuclear matter 
and nuclei has received considerable interest in recent years since it is expected to give
insight
on the partial restoration of chiral symmetry at finite density, on the low-energy dynamics of
QCD, which is related to the $U(1)_A$ anomaly, and on the formation of mesic
bound states with
nuclei--the so called mesic nuclei; see
Refs.~\cite{Khreptak:2023lbh,Bass:2021rch,Haider:2015fea,
Bass:2018xmz,Kelkar:2013lwa} for recent reviews.
Seen the other way around, the discovery of  $\eta$- and $\eta'$-mesic nuclei would allow for a
more accurate determination of the poorly known $\eta N$ and $\eta' N$ interactions;
would open up opportunities to study the structure of these  mesons in a nuclear medium
~\cite{Bass:2005hn}; and provide information on  the dynamics of the axial $U(1)_A$
anomaly in the nuclear  medium~\cite{Li:2022dry}.
 
The concept of mesic nuclei was first introduced by Haider and Liu in Ref.~\cite{Haider:1986sa}.
Mesic nuclei would represent a novel form of nuclear system
in which a meson is bound with a nucleus only through the strong interaction, without the
influence of electromagnetic Coulomb
effects-- such bound state system by the Coulomb interaction
is often called ``mesic atom''.
The $\eta$ and $\eta'$ mesons are particularly promising candidates
for exploring such meson-nucleus bound states~\cite{Metag:2017yuh}.

The experimental searches for this exotic form of nuclear system involve the production
of $\eta$ and $\eta'$ mesons, analyzing their interaction with nuclei, and detecting
$\eta$- and $\eta'$-mesic states through their possible decay modes~\cite{Khreptak:2023lbh,
Bass:2021rch,Haider:2015fea,Bass:2018xmz,Kelkar:2013lwa}.
Despite more than three decades of intense experimental efforts,
unambiguous signals for the $\eta$- and $\eta'$-nucleus bound states
have so far not been directly observed.
However, experimental information on the strength of the $\eta$ and $\eta'$
meson interactions with nuclei has been extracted indirectly from a combination of
measurements (production cross sections,  transparency ratios,
excitation functions and momentum distributions in hadron- and  photo-induced reactions)
and Glauber-, transport-, or  collisional-model
approaches~\cite{Khreptak:2023lbh,Bass:2021rch}.
In this way, the real and imaginary parts of the $\eta$-nucleus optical potential at normal
nuclear density $\rho_0$ have been constrained to $|V_0| \lesssim 60$ MeV and
$|W_0| \lesssim 7 $ MeV, respectively,  using the light nuclei
$^{4}$He~\cite{Skurzok:2018paa,Ikeno:2017xyb}.
Similar analyses have been applied to the $\eta'$-nucleus interaction, giving
$V_0 \approx -40$ MeV~\cite{CBELSATAPS:2016qdi,CBELSATAPS:2013waf,
CBELSATAPS:2018sck} and $W_0 \approx -13$ MeV~\cite{Metag:2017yuh,CBELSATAPS:2012few,
Friedrich:2016cms} using medium and heavy nuclei (C, Nb and Pb).  
A more detailed discussion of the experimental efforts is given in  
Refs.~\cite{Khreptak:2023lbh,Bass:2021rch,Haider:2015fea, Bass:2018xmz,Kelkar:2013lwa}.
Even though in both cases we have an attractive meson-nucleus interaction and a relatively weak
meson absorption, as mentioned above, $\eta$- and $\eta'$-nucleus bound states have so
far not been directly observed~\cite{LEPS2BGOegg:2020cth,Fujioka:2020ewc,
n-PRiMESuper-FRS:2016vbn}, and thus the search for this novel form of
nuclear system continues~\cite{Khreptak:2023lbh,Bass:2021rch,Haider:2015fea,
Bass:2018xmz,Kelkar:2013lwa}.

On the theoretical side, the medium modifications of the $\eta$ and $\eta'$ mesons 
and their nuclear bound states have been studied in a variety of approaches, such as
the quark-meson coupling (QMC) model~\cite{Bass:2005hn, Bass:2013nya},
chiral coupled channels~\cite{Nagahiro:2011fi}, the Nambu--Jona-Lasinio
model~\cite{Bernard:1987sx,Nagahiro:2006dr,Costa:2002gk}, and linear $\sigma$
model~\cite{Sakai:2013nba,Sakai:2022xao},
as well as other approaches~\cite{Nagahiro:2004qz,Jido:2011pq}.
The significant variation of input parameters in different calculations results in a wide 
range  of predicted outcomes. Some of these theoretical analyses predict the
existence of certain $\eta$- and  $\eta'$-mesic nuclei; however, other suggest
these are unlikely;
see Refs.~\cite{Khreptak:2023lbh,Bass:2021rch,Haider:2015fea,Bass:2018xmz,
Kelkar:2013lwa} for a more detailed discussion.
 
In this work we add to the existing theoretical efforts
by updating and extending previous work~\cite{Tsushima:1998qw,Tsushima:1998qp}
on the mass shift of the $\eta$ and $\eta'$ mesons in nuclear matter and their nuclear 
bound states using results obtained  with the  QMC model. The approach followed here is 
an extension of previous work by us in 
Refs.~\cite{Cobos-Martinez:2017vtr,Cobos-Martinez:2017woo,
Cobos-Martinez:2020ynh,Zeminiani:2020aho,Cobos-Martinez:2022fmt}.
Here we use an updated value for the pseudoscalar mixing angle of
$\theta_P=-11.3^{\circ}$ from the Particle Data Group~\cite{Workman:2022ynf}.

This paper is organized as follows. In~\sect{sec:qmc}, we briefly describe
the QMC model and present results for the mass shift of the $\eta$ and
$\eta'$ mesons in symmetric nuclear matter. Using the local density approximation,
in~\sect{sec:nuclear_potential}, we compute potentials for the $\eta$ and $\eta'$ mesons
in nuclei. In~\sect{sec:nuclear_bound_states}, we present numerical results for
the $\eta$- and $\eta'$-nucleus bound state energies by solving
the Schr\"odinger and Klein-Gordon equations with the nuclear potentials
calculated in~\sect{sec:nuclear_potential}.
By adding an imaginary part to the $\eta$-  and $\eta'$-nucleus potentials, to simulate the
absorption of these mesons by nuclei, in~\sect{sec:nuclear_bound_states_complex},
we present our results for the single-particle energies
and absorption widths for the $\eta$ and $\eta'$ in nuclei.
Finally, in \sect{sec:conclusions} we conclude this work.

\section{\label{sec:qmc} Quark-meson coupling model: mass shift}

In a nuclear medium, hadrons with light  quarks, such as the $\eta$ and  $\eta'$ mesons,  are
 expected to change their properties, such as their masses, and thus affect their interaction with 
  nucleons.
  In this section we will compute the mass shift of these mesons in nuclear matter
  using the quark-meson coupling (QMC) model. In later sections,  we will use these results to
  compute the $\eta$-- and $\eta'$--nucleus potentials and single-particle energies.

The QMC model is a quark-based model for nuclear matter and finite nuclei that describes the
 internal structure of the nucleons using the MIT bag model, and the binding of
 nucleons by the self-consistent couplings of the confined light quarks $u$ and $d$ to the 
 scalar-$\sigma$, vector-isoscalar-$\omega$ and vector-isovector-$\rho$ meson fields 
 generated by the confined light quarks in the nucleons~\cite{Guichon:1987jp}.
The QMC model has been successfully applied to investigate the properties of infinite nuclear 
matter and finite nuclei.  
Here we briefly present the necessary details needed to understand our results. For a more in
depth discussion of the model see Refs.~\cite{Saito:2005rv,Guichon:2018uew} and references 
therein.

We consider nuclear matter in its rest frame, where all the scalar and vector mean field potentials,
 which are responsible for the nuclear many-body interactions, are constants in Hartree
 approximation.
 Assuming $SU(2)$ symmetry for the quarks ($m_q = m_u = m_d$ and $q = u$ or $d$ below)
the Dirac equations for the quarks and antiquarks in nuclear matter, in a hadron bag 
$h=\eta,\,\eta'$ at the position $x=(t,\vec{r})$ (with $|\vec{r}| \le R_h^{*}$, $R_h^{*}$ being the 
in-medium bag radius), neglecting the Coulomb force, are given by~\cite{Guichon:1987jp,
Saito:2005rv,Guichon:2018uew,Tsushima:1997df}:

\begin{eqnarray}
&&\hspace{-6ex}\left[i\cancel{\partial}_{x} - \left(m_{q} - V^{q}_{\sigma}\right)
\mp \gamma^{0}\left(V^{q}_{\omega} + \frac{1}{2}V^{q}_{\rho} \right) \right]
\begin{pmatrix}
        \psi_{u}\left(x\right)\\
        \psi_{\overline{u}}\left(x\right)
       \end{pmatrix} = 0, \\
&&\hspace{-6ex}\left[i\cancel{\partial}_{x} - \left(m_{q} - V^{q}_{\sigma}\right)
\mp \gamma^{0}\left(V^{q}_{\omega} - \frac{1}{2}V^{q}_{\rho} \right)\right]
\begin{pmatrix}
        \psi_{d}\left(x\right)\\
        \psi_{\overline{d}}\left(x\right)
       \end{pmatrix} = 0, \\
&&\left[i\gamma \cdot \partial_{x} - m_{s}\right]\psi_{s, \overline{s}}\left(x\right) = 0.
\end{eqnarray}
Here the (constant) mean-field potentials for the light quark $q$ in nuclear matter are defined by 
$V^{q}_{\sigma} \equiv g^{q}_{\sigma}\sigma$,  
$V^{q}_{\omega} \equiv g^{q}_{\omega}\omega = g^q_\omega\, \delta^{\mu 0} \omega^\mu$, 
$V^{q}_{\rho} \equiv g^{q}_{\rho}b = g^q_\rho\, \delta^{i3} \delta^{\mu 0} \rho^{i,\mu}$, 
with the $g^{q}_{\sigma}$, $g^{q}_{\omega}$ and $g^{q}_{\rho}$ being the corresponding 
quark-meson coupling constants.
Note that $V_{\rho}^{q} \propto (\rho_p - \rho_n) = 0$ in symmetric nuclear matter, although this is
not true in a nucleus where the Coulomb force may induce an asymmetry between the proton and
 neutron distributions even in a nucleus with the same number of protons and neutrons, resulting 
 in $V_{\rho}^{q} \ne 0$ at a given position in a nucleus.
 The static solution for the ground state quarks (antiquarks) with a flavor $f (=u,d,s)$ is written 
 as $\psi_{f}\left(x\right) = N_{f}e^{-i\epsilon_{f}t/R^{*}_{h}}\psi_{f}\left(\textbf{r}\right)$, 
with the $N_f$ being the normalization factor, and $\psi_{f}\left(\textbf{r}\right)$ the corresponding 
spin and spatial part of the wave function.
The eigenenergies for the quarks and antiquarks in the $\eta$ and $\eta'$ mesons, in units of 
$1/R^{*}_{\eta,\eta'}$, are given by:
\begin{eqnarray}
&&\begin{pmatrix}
        \epsilon_{u}\\
        \epsilon_{\overline{u}}
       \end{pmatrix} = \Omega^{*}_{q} \pm R^{*}_{\eta,\eta'} 
\left(V^{q}_{\omega} + \frac{1}{2}V^{q}_{\rho}\right),  \\ 
&&\begin{pmatrix}
        \epsilon_{d}\\
        \epsilon_{\overline{d}}
       \end{pmatrix} = \Omega^{*}_{q} \pm R^{*}_{\eta,\eta'} 
\left(V^{q}_{\omega} - \frac{1}{2}V^{q}_{\rho}\right),\\
&&\epsilon_{s} = \epsilon_{\overline{s}} = \Omega_{s}.
\end{eqnarray}
The in-medium mass $m^{*}_h$ and bag radius $R^{*}_h$ of hadron $h$ are determined from
\begin{eqnarray}
\label{eqn:mhnm}
m_\eta^*
&=& \frac{2 [a_{P}^2\Omega_q^* + b_{P}^2\Omega_s] - z_\eta}{R_\eta^*}
+ {4\over 3}\pi R_\eta^{* 3} B, \\
& &\hspace{-5ex}({\rm for}\hspace{1ex} \eta',\hspace{1ex}
\eta \to \eta',\, {\rm and}\hspace{1ex}
a_P \leftrightarrow b_P),
\nonumber\\
& & \label{eqn:stab}
\left.\frac{d m_h^*}
{d R_j}\right|_{R_h = R_h^*} = 0, \quad\quad
(h =\eta,\eta'),
\\
a_{P} &\equiv& \sqrt{1/3} \cos\theta_{P}
- \sqrt{2/3} \sin\theta_{P}, \\
b_{P} & \equiv & \sqrt{2/3} \cos\theta_{P}
+ \sqrt{1/3} \sin\theta_{P},\qquad
\label{abpv}
\end{eqnarray}
where $\Omega^{*}_{q} = \Omega^{*}_{\overline{q}} = \left[x^{2}_{q} + \left(R^{*}_{\eta,\eta'} 
m^{*}_{q}\right)^{2}\right]^{1/2}$, and $m^{*}_{q} = m_{q} - g^{q}_{\sigma}\sigma$
and $\Omega^{*}_{s} = \Omega^{*}_{\overline{s}} = \left[x^{2}_{s} + \left(R^{*}_{\eta,\eta'} 
m_{s}\right)^{2}\right]^{1/2}$, with $x_{q,s}$ being the lowest mode bag eigenfrequencies. 
$B$ is the bag constant; $n_{q,s}$ ($n_{\overline{q},\overline{s}}$) are the lowest 
mode valence quark (antiquark) numbers  for the quark flavors $q$ and $s$ 
in the corresponding $\eta$ and $\eta'$ mesons;  and $z_{\eta,\eta'}$ parameterize the sum 
of the center-of-mass  and gluon fluctuation effects and are 
assumed to be independent of density~\cite{Guichon:1995ue}. The MIT big parameters $z_N$ 
($z_h$) and $B$ are fixed by fitting the nucleon (hadron) mass in free space.

We choose the values ($m_q, m_s$) = (5, 250) MeV for the current quark masses, and $R_{N}$ 
= 0.8 fm for the free space nucleon bag radius.
(See Ref.~\cite{Tsushima:2020gun} for ($m_q, m_s$) = (5, 93) MeV result.)
The quark-meson coupling constants,
$g^{q}_{\sigma}$, $g^{q}_{\omega}$ and $g^{q}_{\rho}$ used for the light quarks   
in the $\eta$ and $\eta'$ mesons (the same as in the nucleon), were determined by the fit 
to the saturation energy (-15.7 MeV) at the saturation density  
($\rho_0 = 0.15$ fm$^{-3}$) of symmetric nuclear matter for $g^q_\sigma$ and 
$g^q_\omega$, and by the bulk symmetry energy (35 MeV) for
$g^q_\rho$~\cite{Guichon:1987jp,Saito:2005rv}.
The obtained values for the quark-meson coupling constants are 
($g_\sigma^q$, $g_\omega^q$, $g_\rho^q$)= (5.69, 2.72, 9.33).

Finally, for the mixing angle $\theta_P$ we use the value $\theta_P=-11.3^{\circ}$, neglecting any possible mass dependence and imaginary parts~\cite{Workman:2022ynf,Tsushima:2020gun}.
Furthermore, we also assume that the value of the mixing angle does not change in the nuclear medium.
\begin{figure}
\centering
 \includegraphics[scale=0.3]{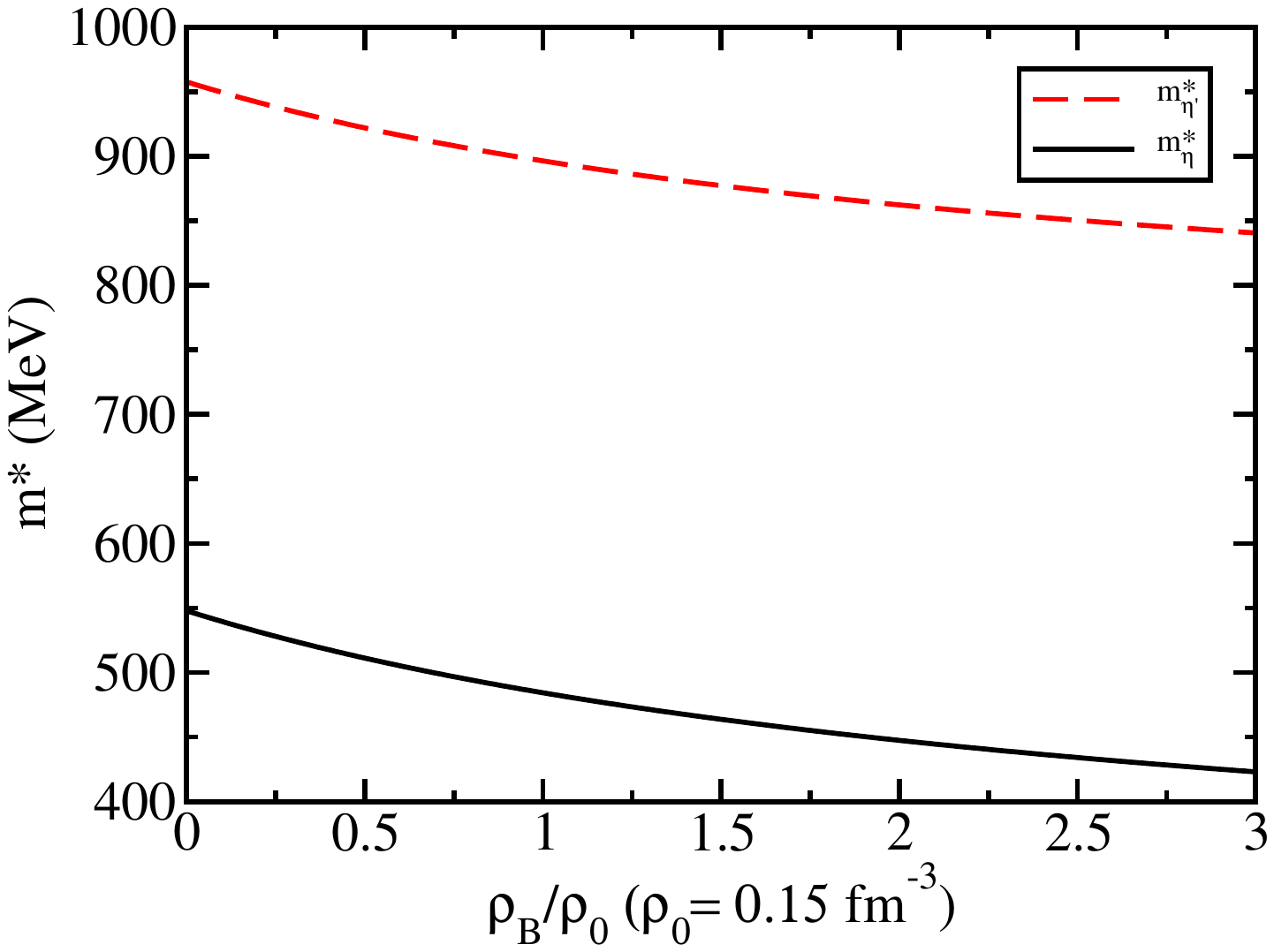} 
 \includegraphics[scale=0.3]{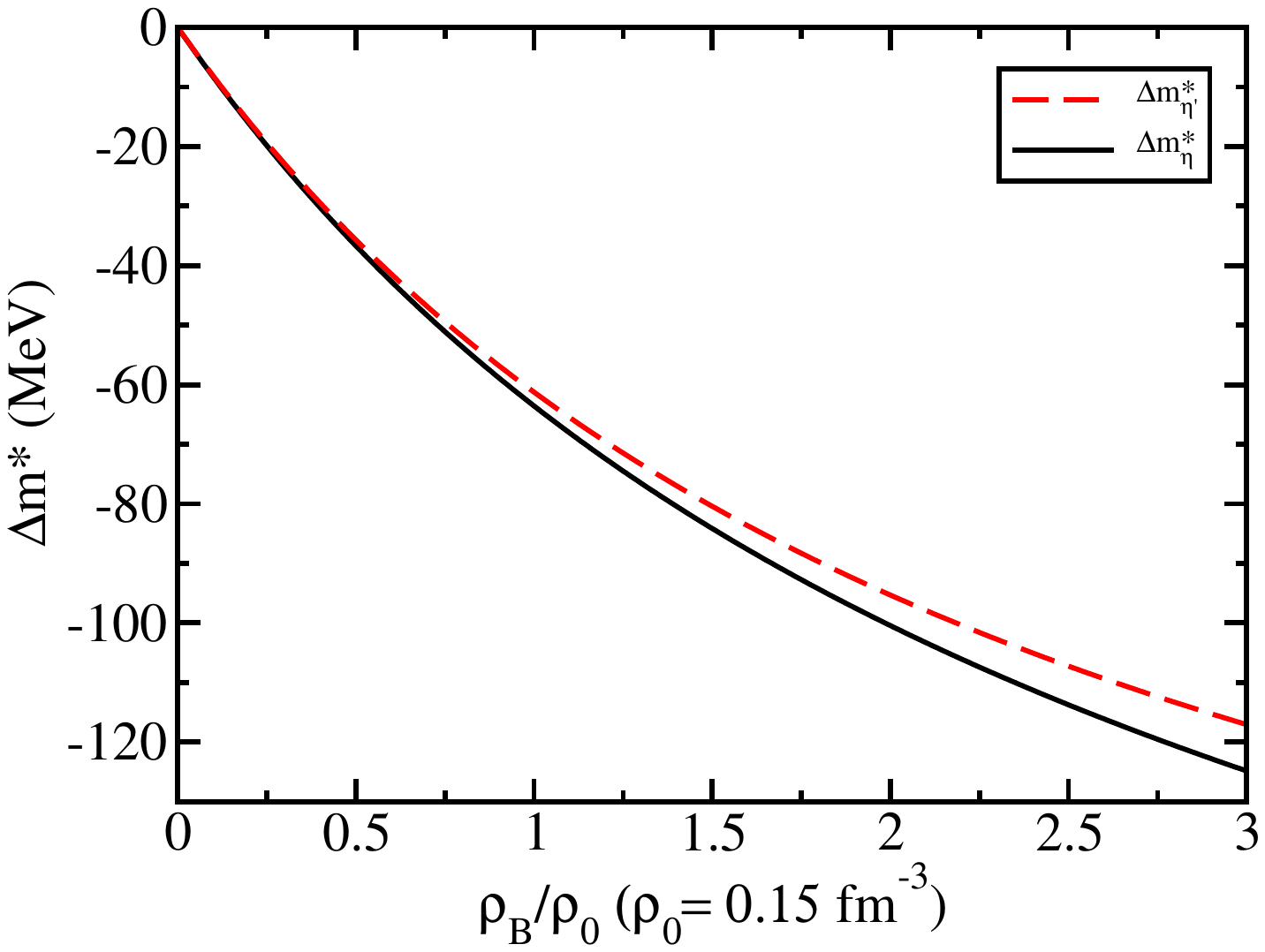}
  \caption{\label{fig:metaetaprime-nm} $\eta$ and $\eta'$ masses (top panel)
  and mass shift
  (bottom panel) in nuclear matter calculated in the quark-meson coupling (QMC) model as a 
  function of the nuclear matter density $\rho_B$.  }
\end{figure}

In \fig{fig:metaetaprime-nm}, present the QMC model predictions for the masses and
mass shift $\Delta m_h(\rho_B)\equiv m_h^*(\rho_B)-m_h$, where $m_h^*$ is the
in-medium meson mass and $m_h$ is vacuum mass, for the $\eta$ and  $\eta'$ mesons 
in symmetric nuclear matter as a function of the nuclear matter density $\rho_B$. 
Clearly, the masses of these meson decrease in the nuclear
medium, a clear signature of the partial restoration of chiral symmetry in medium.
The extracted values for the mass shift at nuclear matter saturation density is
$\Delta m_{\eta}(\rho_0)=-63.6$ MeV and $\Delta m_{\eta'}(\rho_0)=-61.3$ MeV,
respectively.
These results are consistent  with the values  extracted for these quatities from experimental 
data, using transport-, collision-, or Glauber calculations of $V_0= -60$ to 0 MeV
for the $\eta$ meson  and $-40\pm 6 \pm 15$ MeV  for the $\eta'$ 
meson~\cite{Bass:2021rch,Khreptak:2023lbh}.
\begin{figure}
\centering
 \includegraphics[scale=0.3]{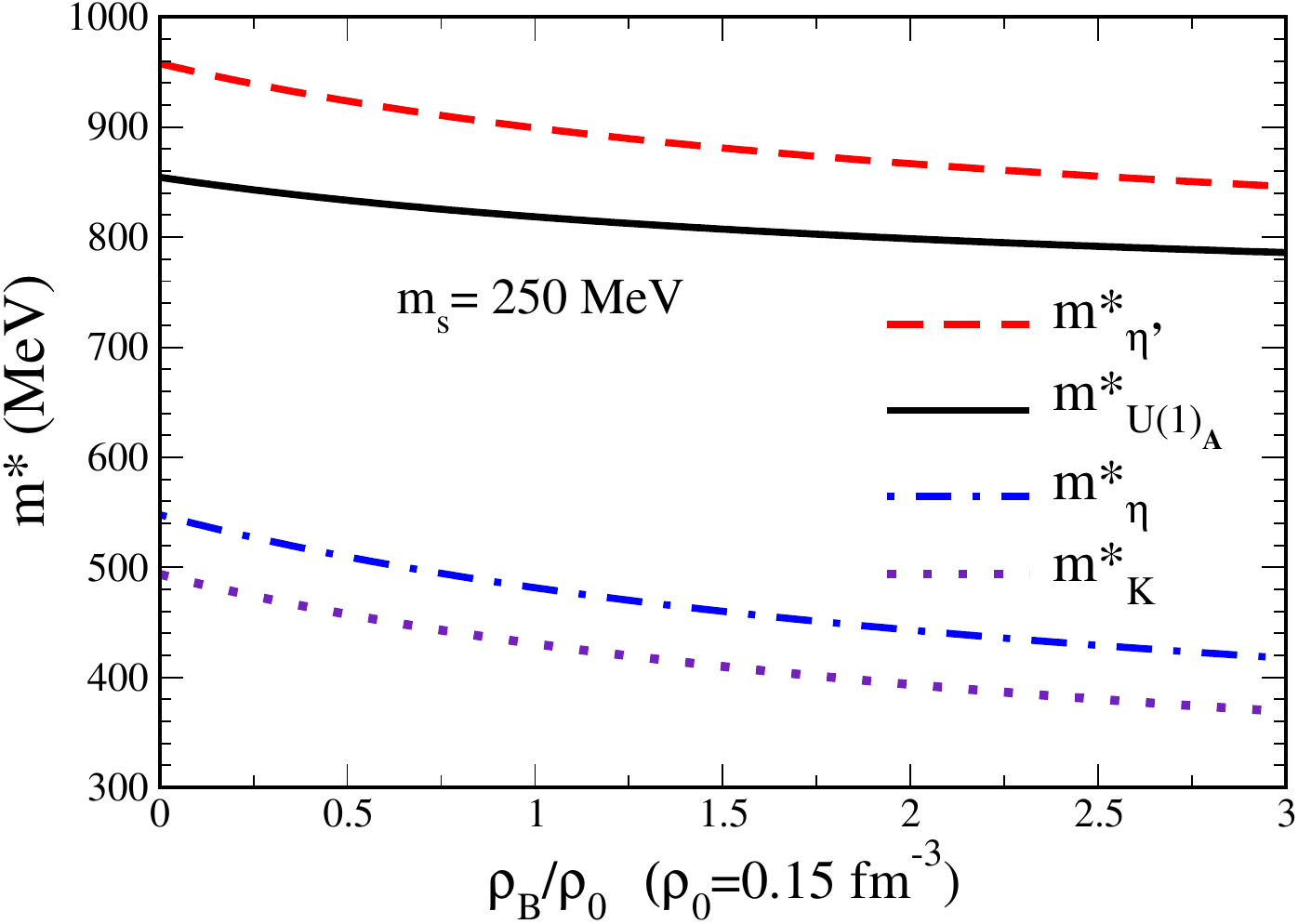} 
  \caption{\label{fig:U1A_mass} $U(1)_{A}$ mass shift in nuclear matter calculated in the 
  quark meson coupling (QMC) model as a  function of the nuclear matter density $\rho_B$.}
\end{figure}

Finally, in \fig{fig:U1A_mass} we present results, for the first time, for the
$U(1)_A$ mass shift $M_{U(1)_A} \equiv  m_{\eta'}^{2} + m_{\eta}^2 -2m_{K}^{2}$,
where $m_{K}$
is the kaon mass, in nuclear matter as a  function of the nuclear matter density $\rho_B$.
The $U(1)_A$ mass shift  is  related to the topological susceptibility 
and parametrizes the deviation from the $U(1)_A$ symmetric world~\cite{Feldmann:1999uf}.
Since $M_{U(1)_A}=0$ means a $U(1)_A$ symmetric world, this result implies an effective
partial restoration of the $U(1)_A$ symmetry  in nuclear matter. However, a more
complete study of the $U(1)_A$ symmetry in nuclear matter, in the approach followed in this
work, deserves further investigation. At the moment, there is no experimental information
on the possible effective restoration of the $U(1)_A$ symmetry at finite density.

\section{\label{sec:nuclear_potential} $\eta$-- and $\eta'$--nucleus potentials}

The mass shift for the $\eta$ and $\eta'$ in nuclear matter was calculated in the
previous section as a function the nuclear matter density $\rho_B$,
using the quark-meson coupling (QMC) model, see~\fig{fig:metaetaprime-nm}.
These results show that the nuclear medium provides attraction to
 these mesons and open the possibility to study the binding of theses mesons to nuclei, 
 which we carry out in this section. First, we obtain the $\eta$ and $\eta'$
Lorentz scalar potentials in nuclei and then we solve the Schr\"{o}dinger and 
Klein-Gordon equations for these mesons using these potentials for various nuclei to 
obtain the $\eta$ and $\eta'$ single-particle energies of the $\eta$- and $\eta'$-nuclear 
bound states. We consider the situation in which these mesons have been produced 
nearly at rest inside a nucleus $A$, and study the following
nuclei in a wide range of masses, namely $^{4}_{2}$He, $^{12}_{6}$C, $^{16}_{8}$O,
$^{40}_{20}$Ca, $^{48}_{20}$Ca, $^{90}_{42}$Zr, and $^{208}_{82}$Pb.

The $\eta$ and $\eta'$ Lorentz scalar potentials in nuclei (no Lorentz vector potentials 
due to the $q\bar{q}$ structure) are calculated in the local density approximation
\begin{equation}
\label{eqn:Vs}
V_{h A}(r)= \Delta m_{h}(\rho^{A}_{B}(r)),
\end{equation}
\noindent where  $\Delta m_{h}(\rho_B)= m_{h}^{*}(\rho^{A}_{B}(r))-m_{h}$ 
($h=\eta,\,\eta'$)  is the mass shift in nuclear matter as a function of the nuclear matter density
$\rho_B$; $\rho^{A}_{B}(r)$ is the baryon density distribution of nucleons
for {a nucleus $A$;
$r$ is the distance from the center of the nucleus and $m_h$ is the mass of the meson 
in vacuum ($m_\eta= 547.862$ MeV and $m_{\eta'}=957.78$ MeV)

The  baryon density distributions for the nuclei listed above, that enter
into~\eqn{eqn:Vs}, are calculated in the QMC model~\cite{Saito:1996sf},
except for $^4$He, which was parameterized in Ref.~\cite{Saito:1997ae}.
\begin{figure}[ht]
\centering
\scalebox{0.9}{
\begin{tabular}{cc}
 \includegraphics[scale=0.3]{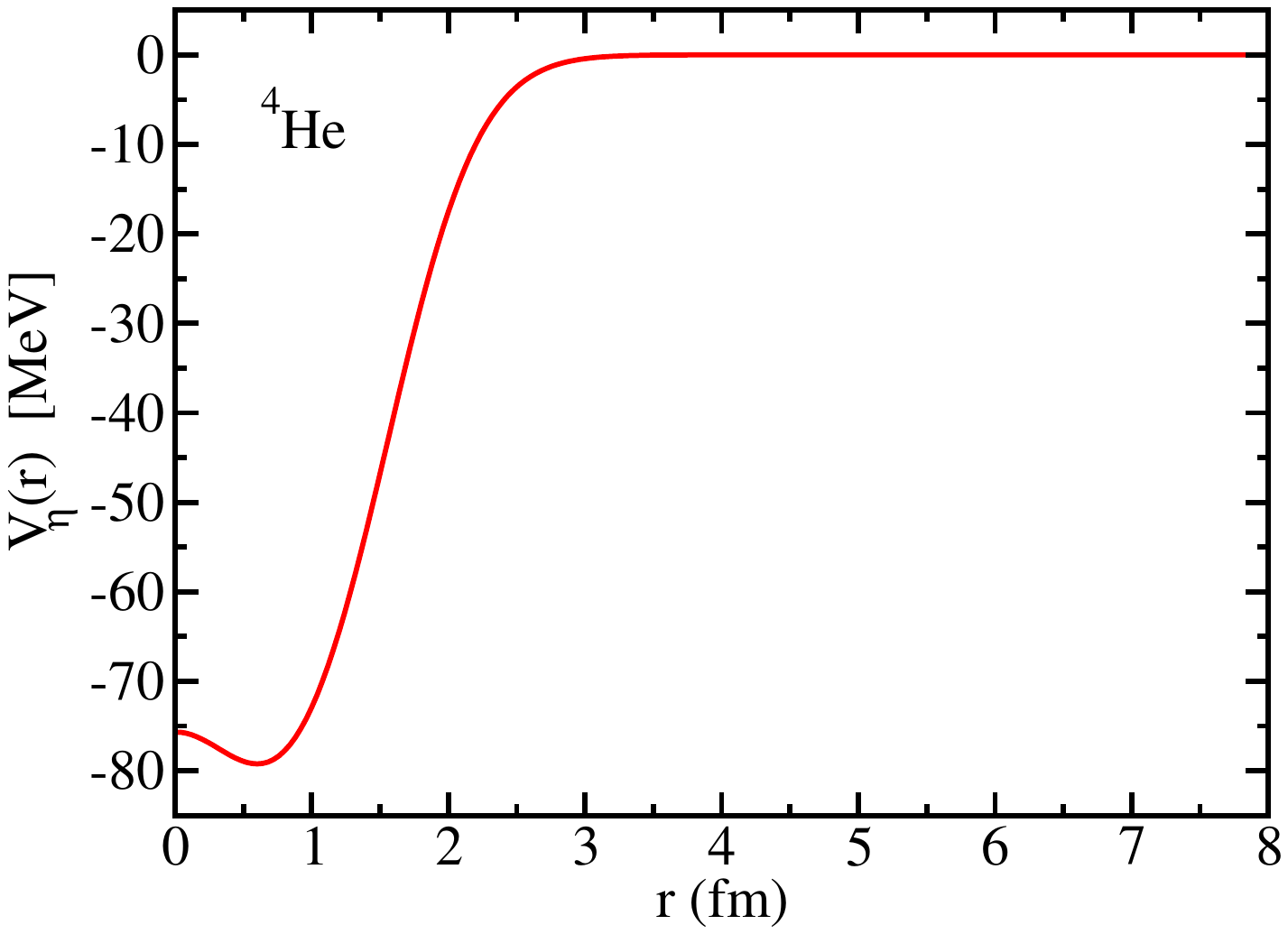} &
 \includegraphics[scale=0.3]{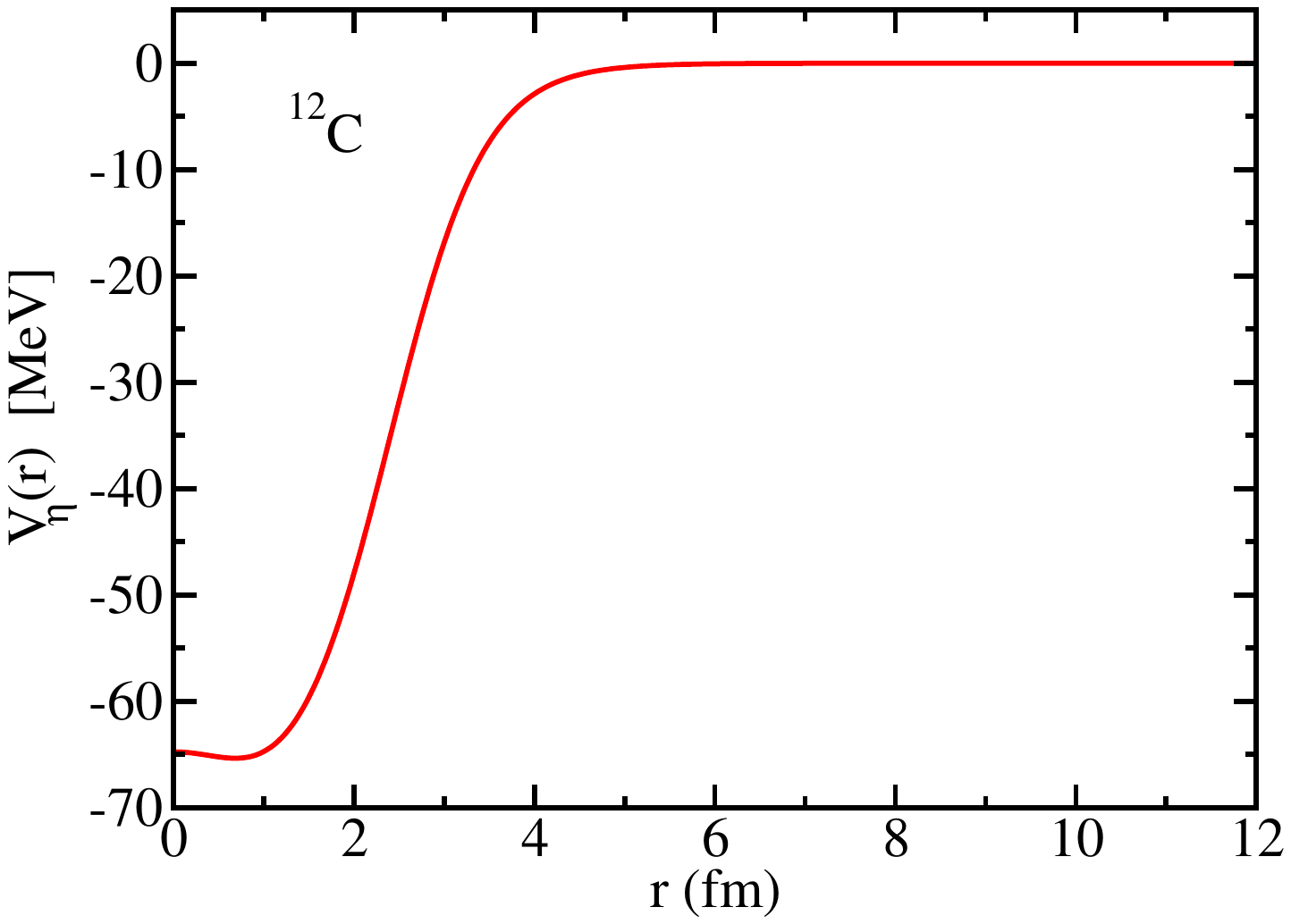} \\
 \includegraphics[scale=0.3]{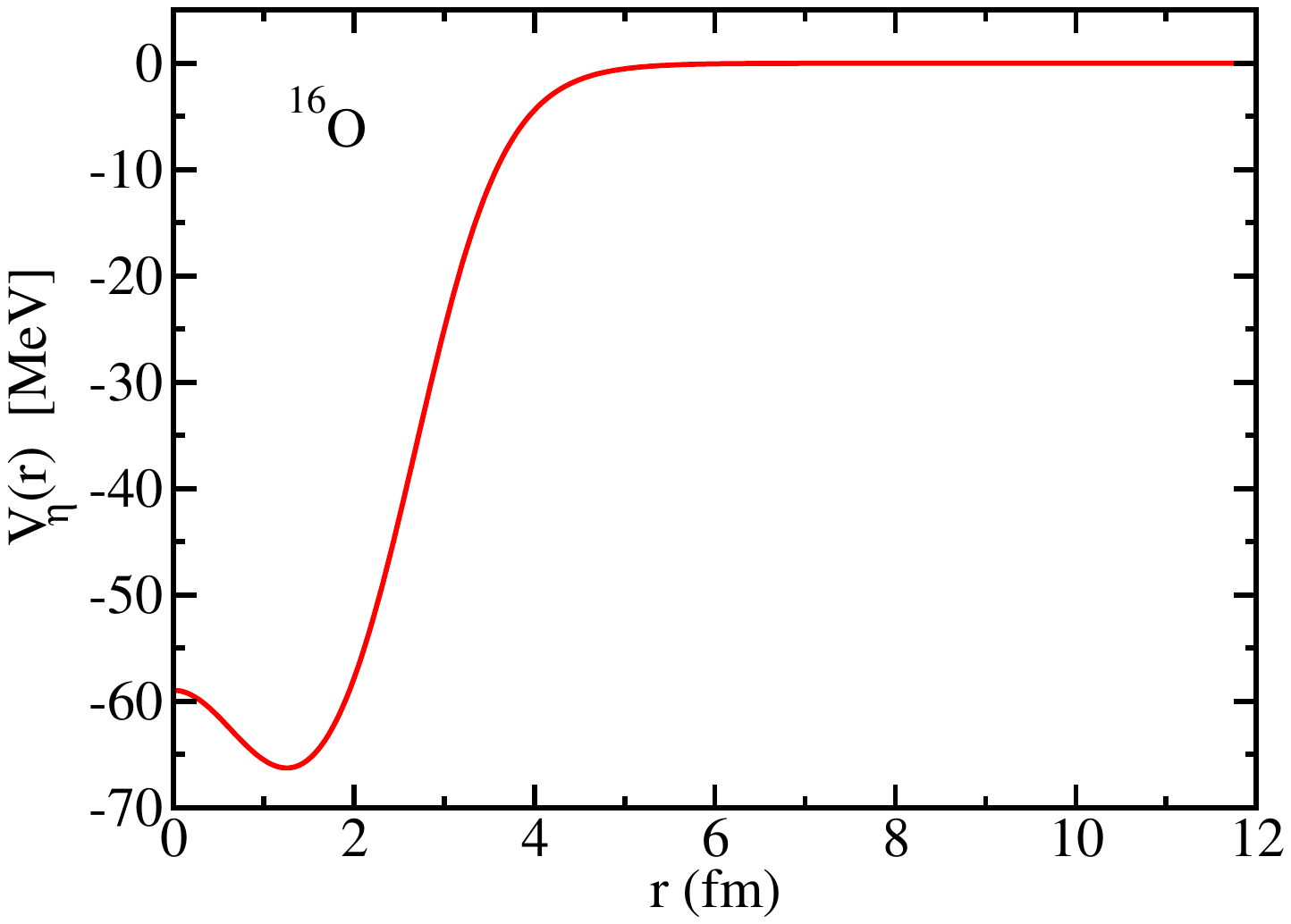} &
 \includegraphics[scale=0.3]{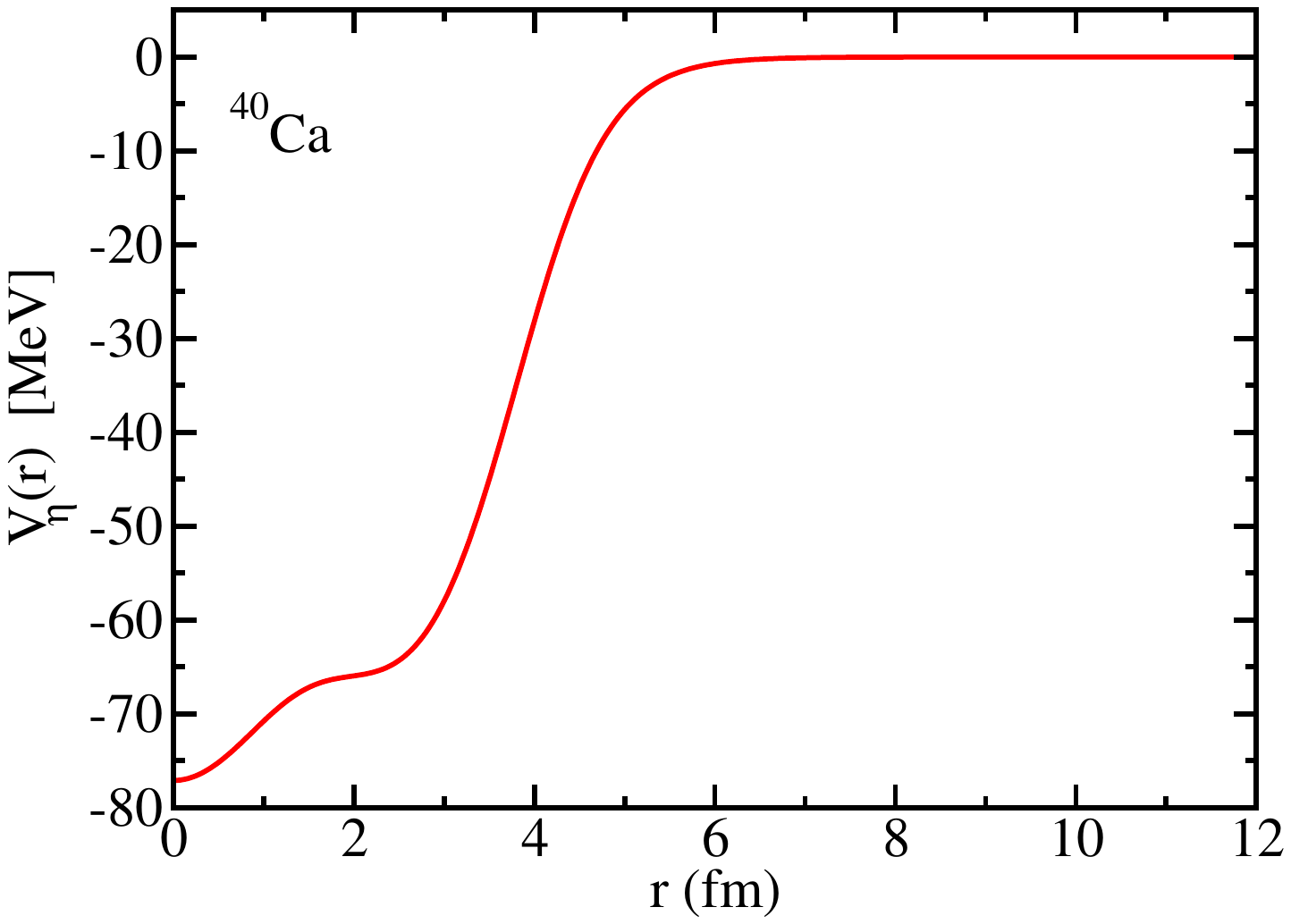} \\
 \includegraphics[scale=0.3]{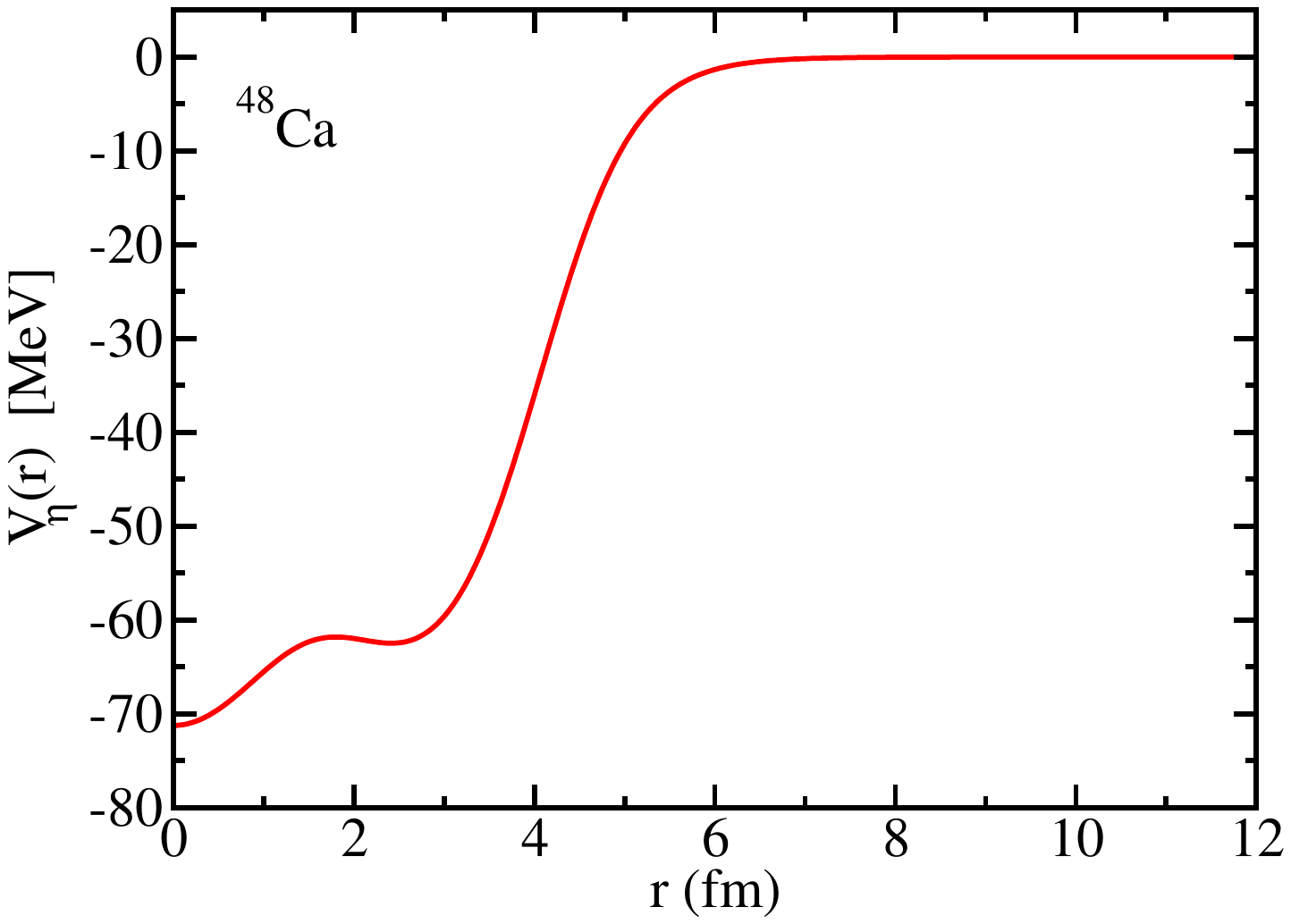} &
 \includegraphics[scale=0.3]{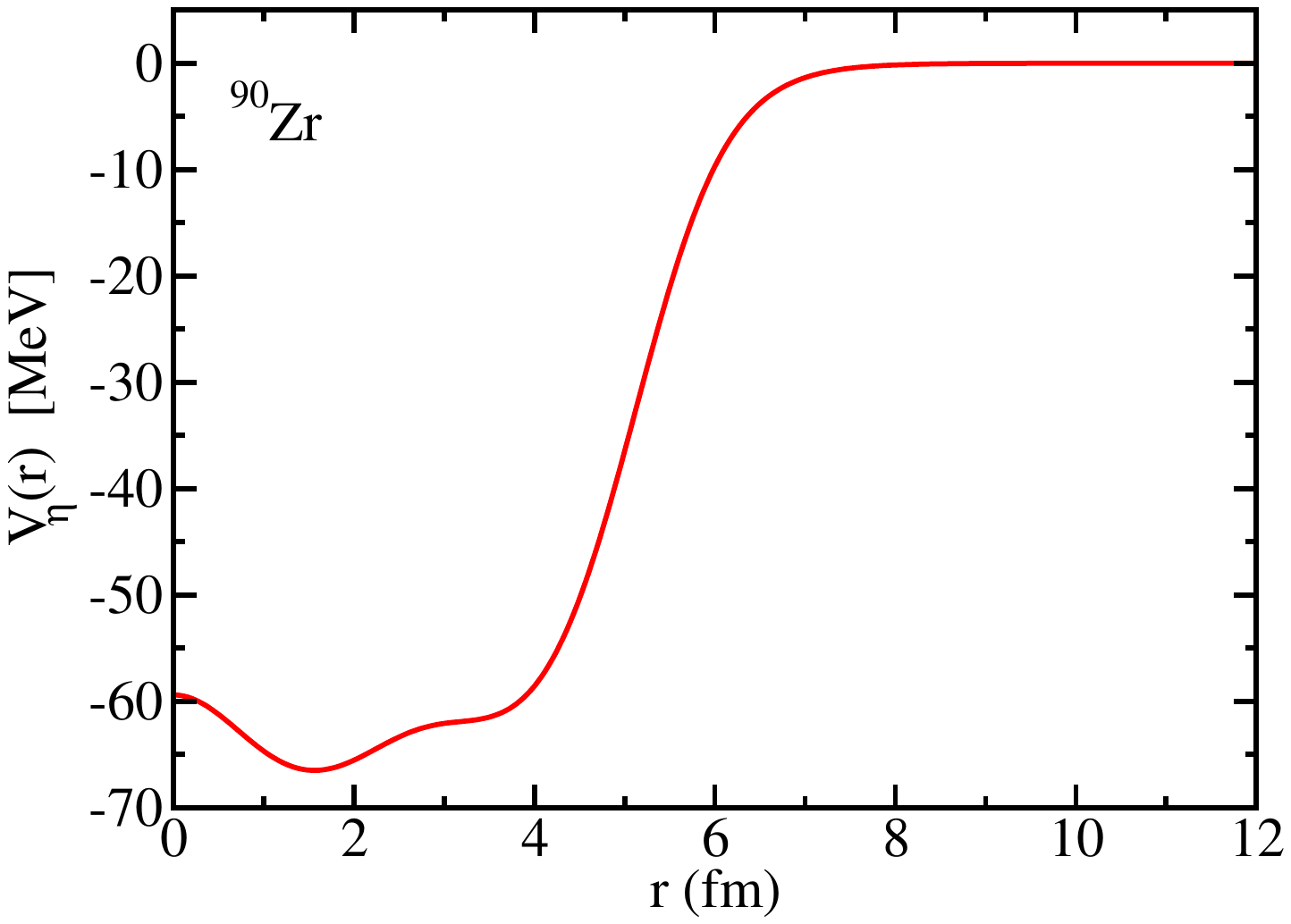} \\
 \includegraphics[scale=0.3]{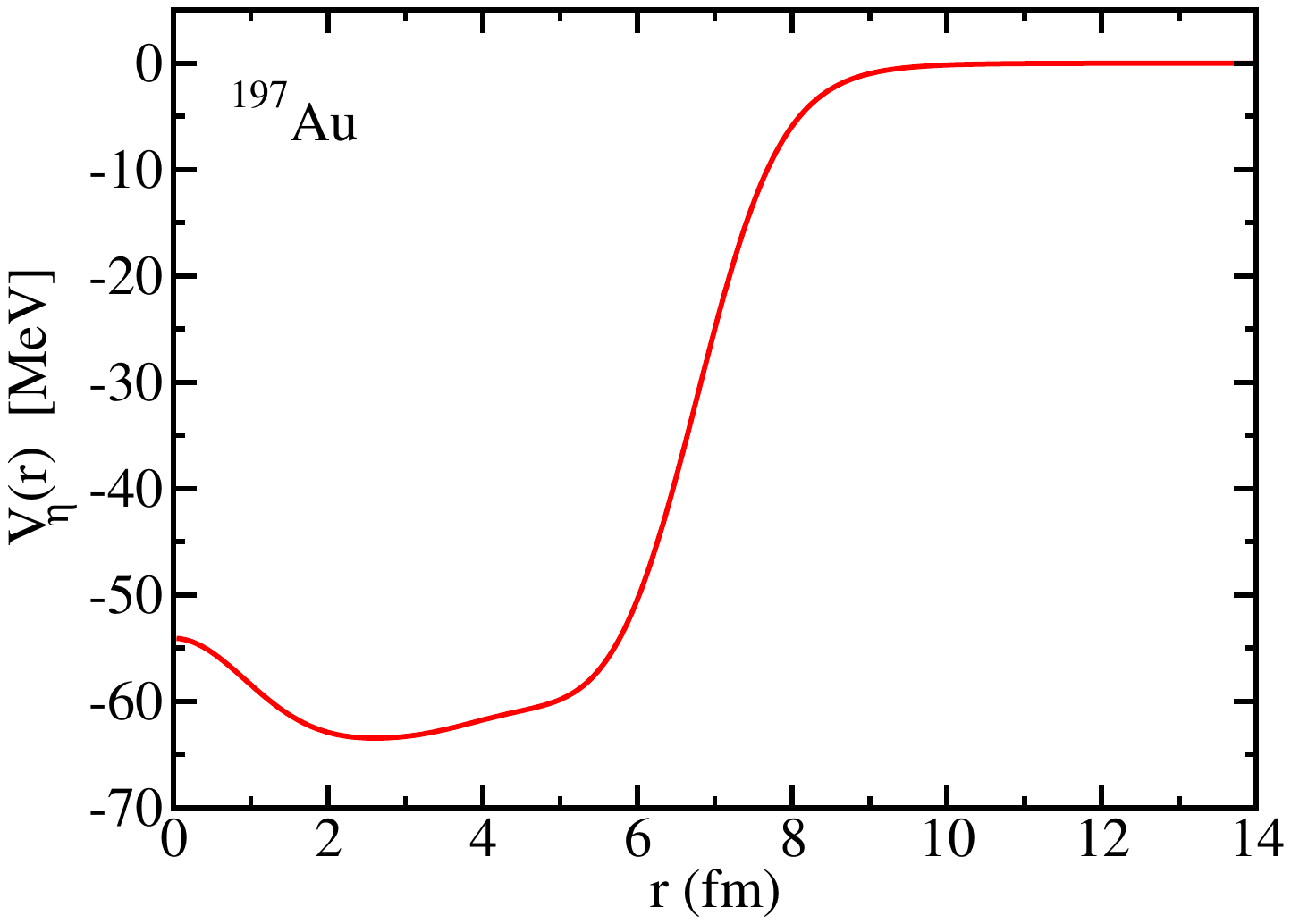}  &
 \includegraphics[scale=0.3]{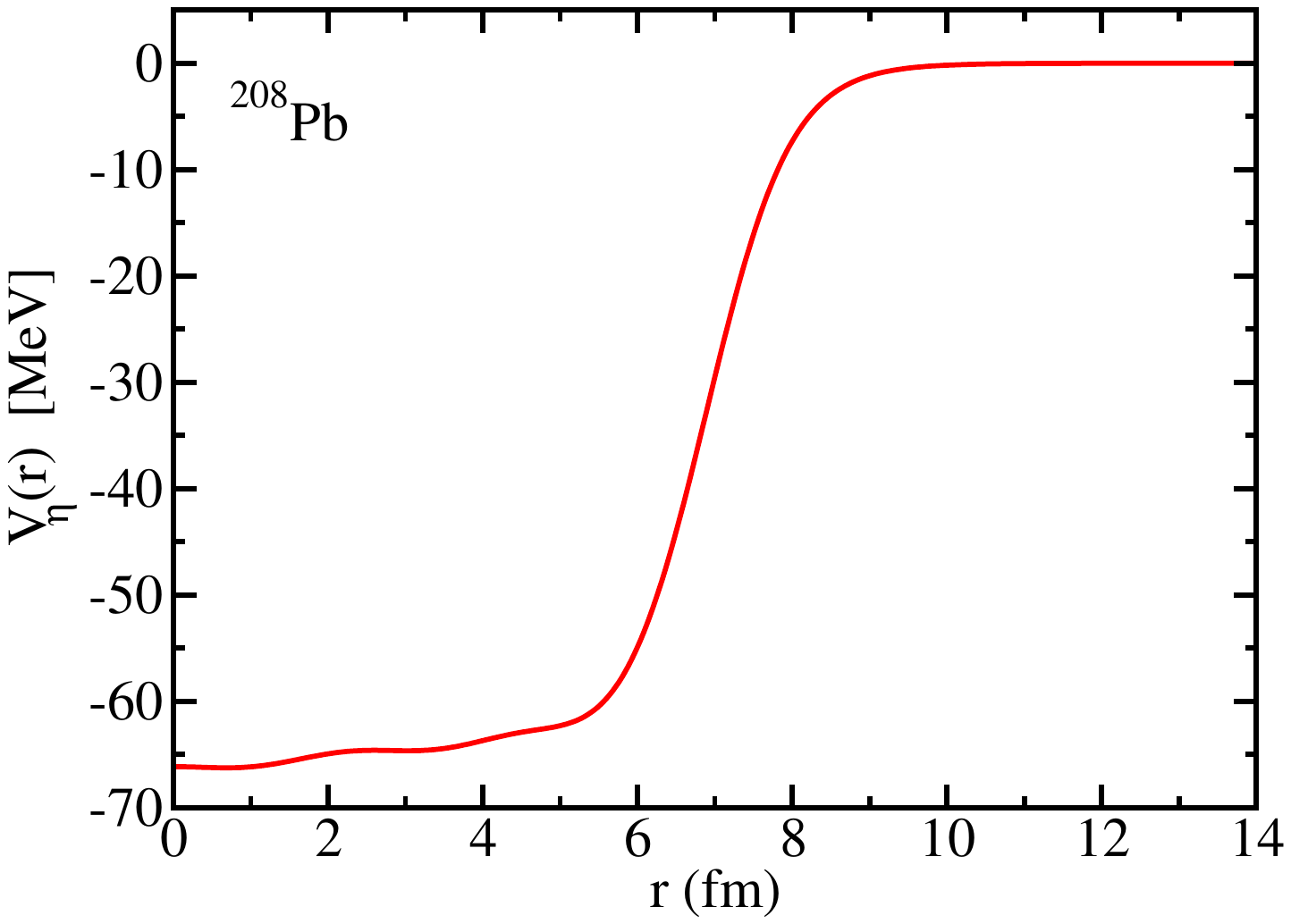} 
 \end{tabular}
 }
\caption{\label{fig:v_eta_nucl} $\eta$ scalar potentials for several nuclei.}
\end{figure}
The calculated potentials for the $\eta$ and $\eta'$ mesons in nuclei are shown in
Figs.~\ref{fig:v_eta_nucl} and~\ref{fig:v_etaprime_nucl}.
These figures show that all potentials for the  $\eta$ and $\eta'$ in nuclei are attractive. 
This is so because the corresponding mass shift (in nuclear matter) is negative for both mesons.
The differences in the potentials, for a given meson, reflect the differences in the baryon
density distributions for the nuclei studied.
Furthermore, note that for a given nucleus, the potentials for
the $\eta$ and $\eta'$ are very similar, the reason for this
being that $\to$ similarity is becausethe mass shift for the
$\eta$ and $\eta'$ is very similar, as shown in \fig{fig:metaetaprime-nm}.

\begin{figure}[ht]
\centering
\scalebox{0.9}{
\begin{tabular}{cc}
 \includegraphics[scale=0.3]{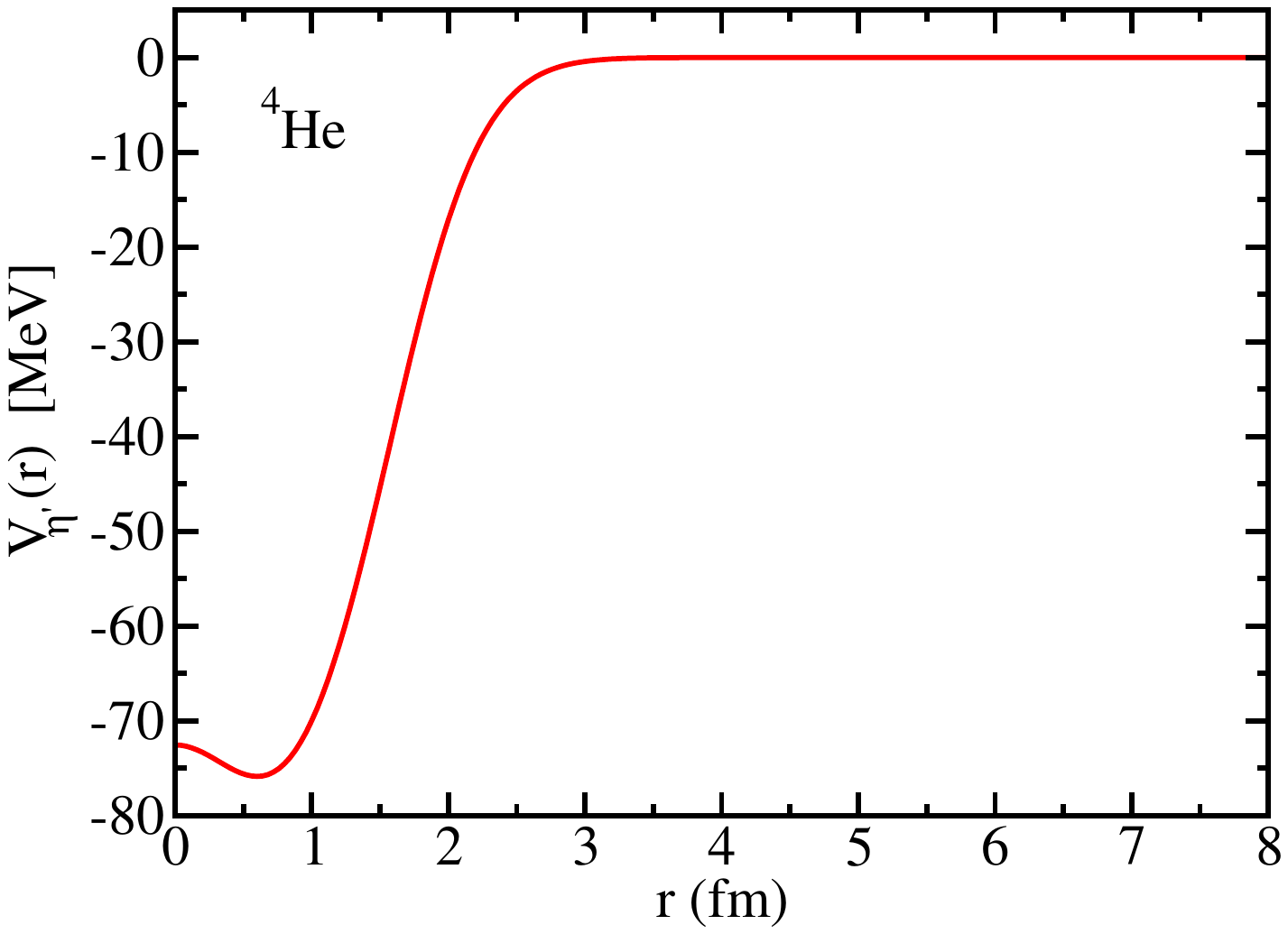} &
 \includegraphics[scale=0.3]{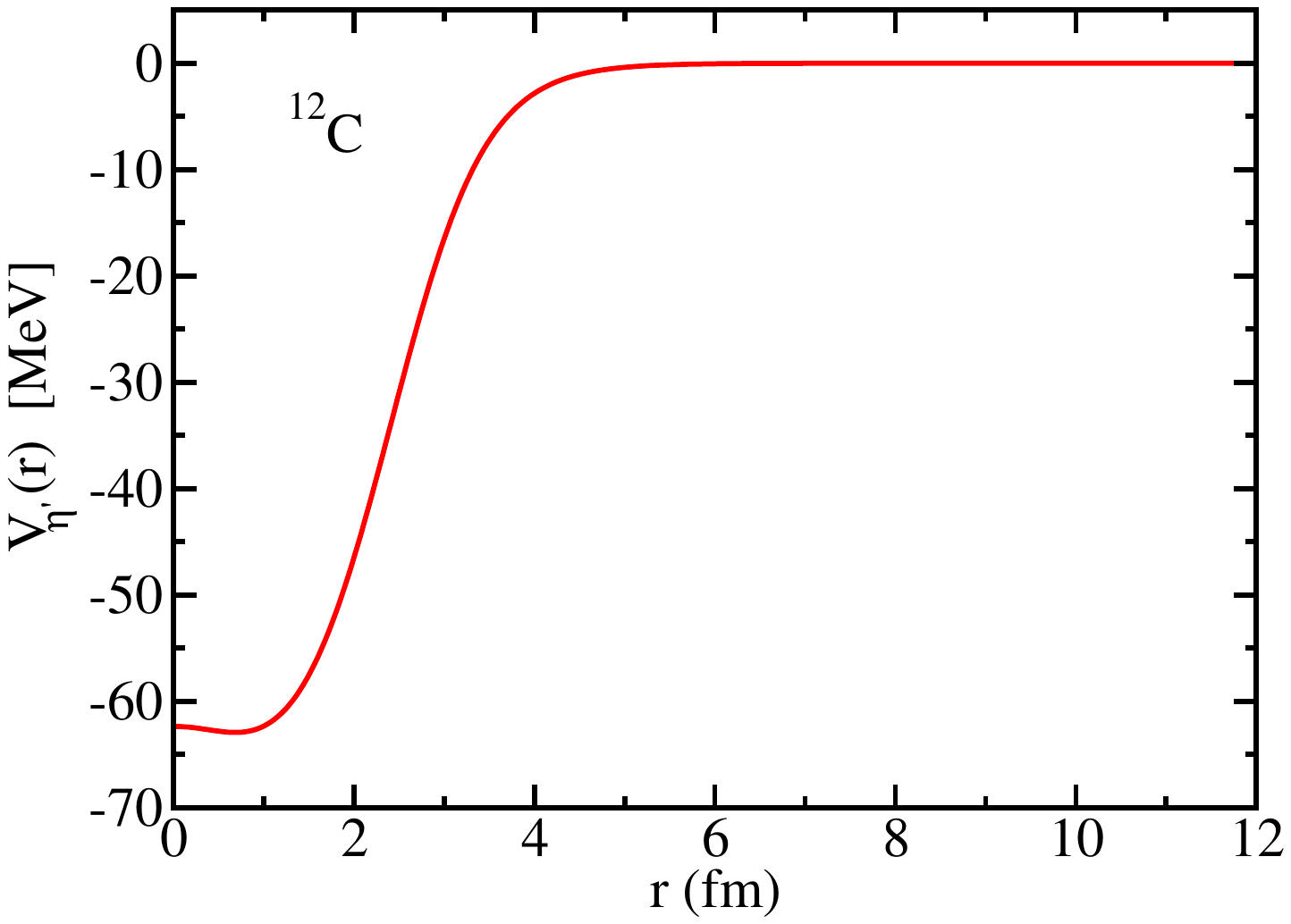} \\
 \includegraphics[scale=0.3]{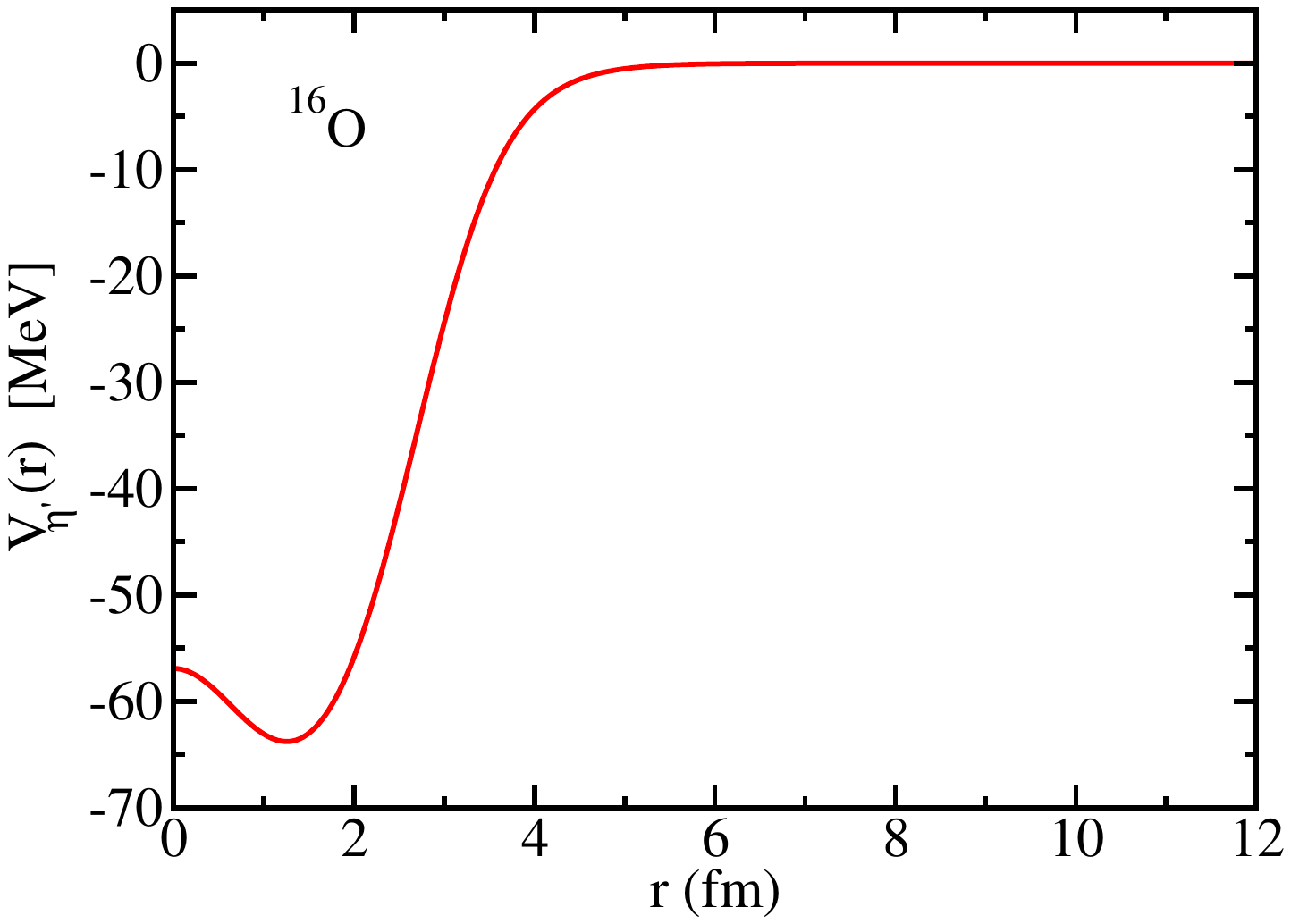} &
 \includegraphics[scale=0.3]{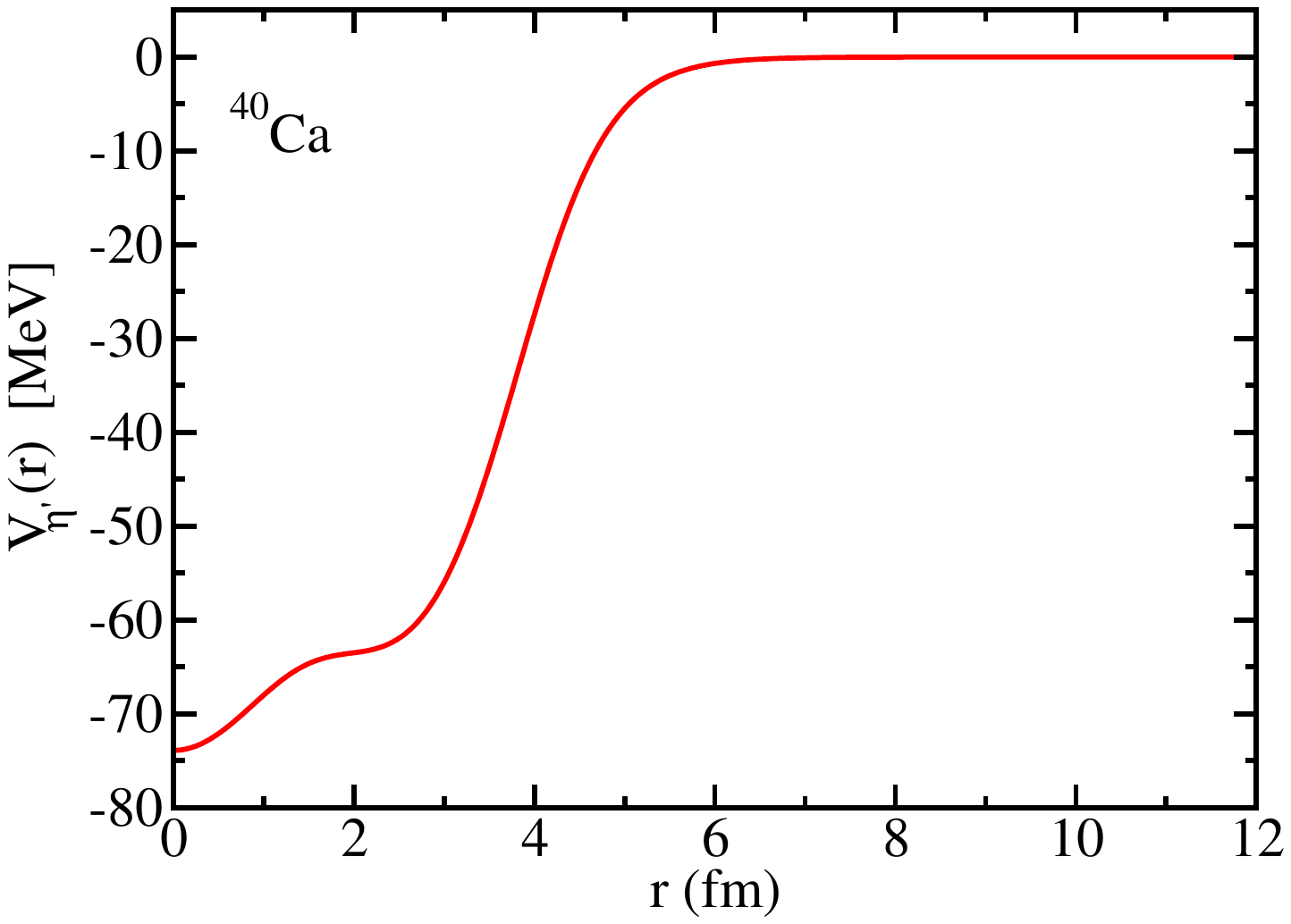} \\
 \includegraphics[scale=0.3]{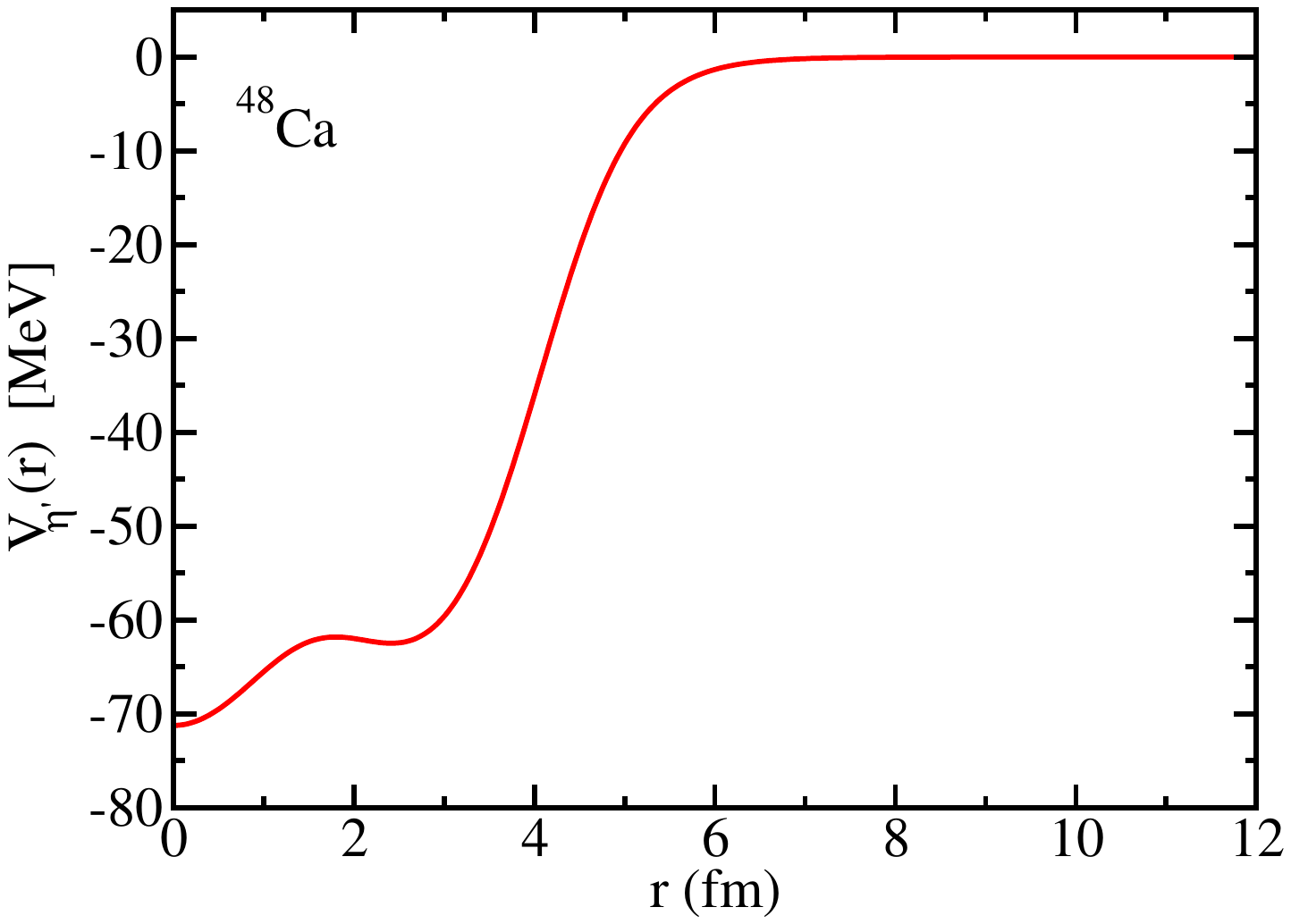} &
 \includegraphics[scale=0.3]{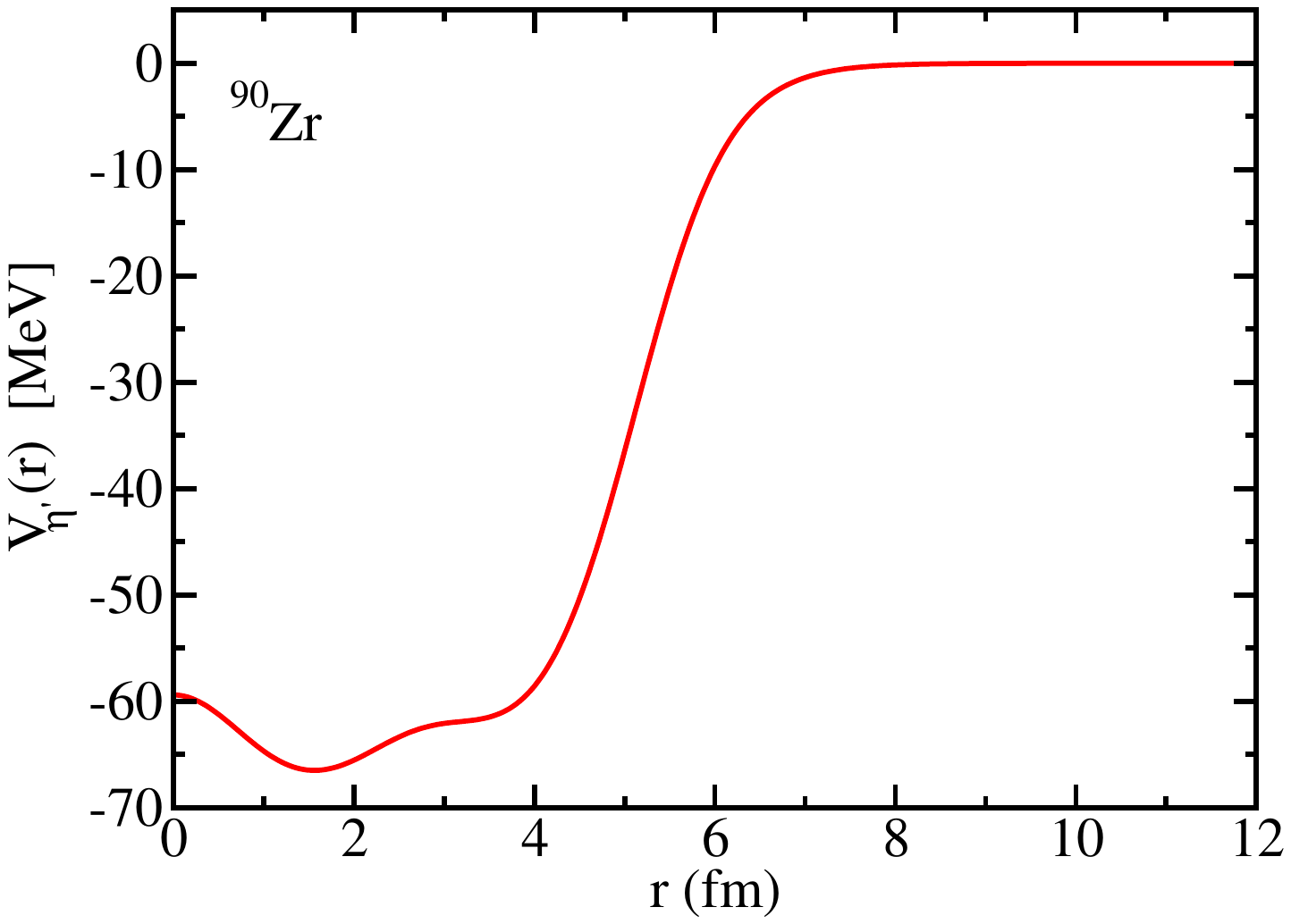} \\
 \includegraphics[scale=0.3]{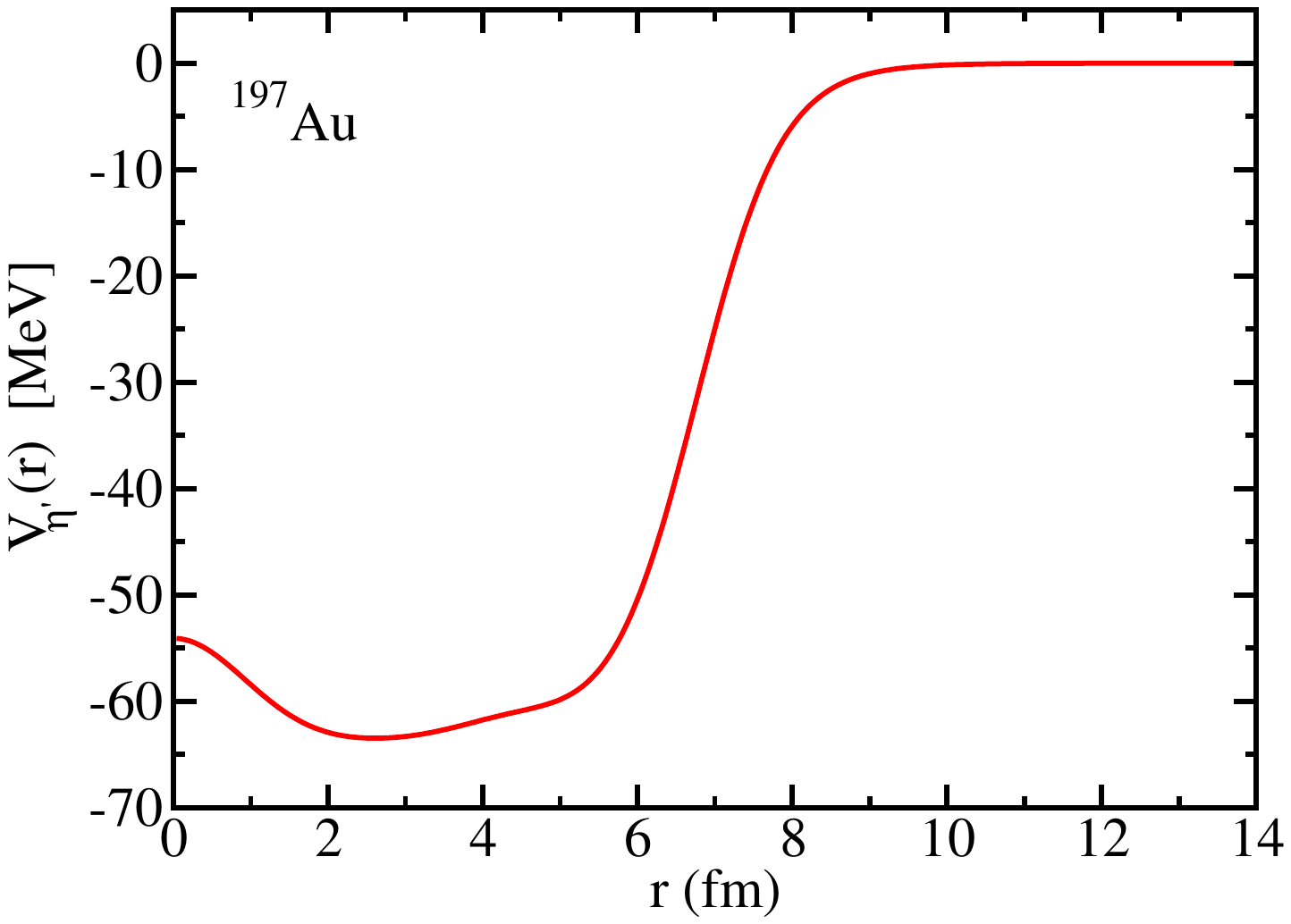}  &
 \includegraphics[scale=0.3]{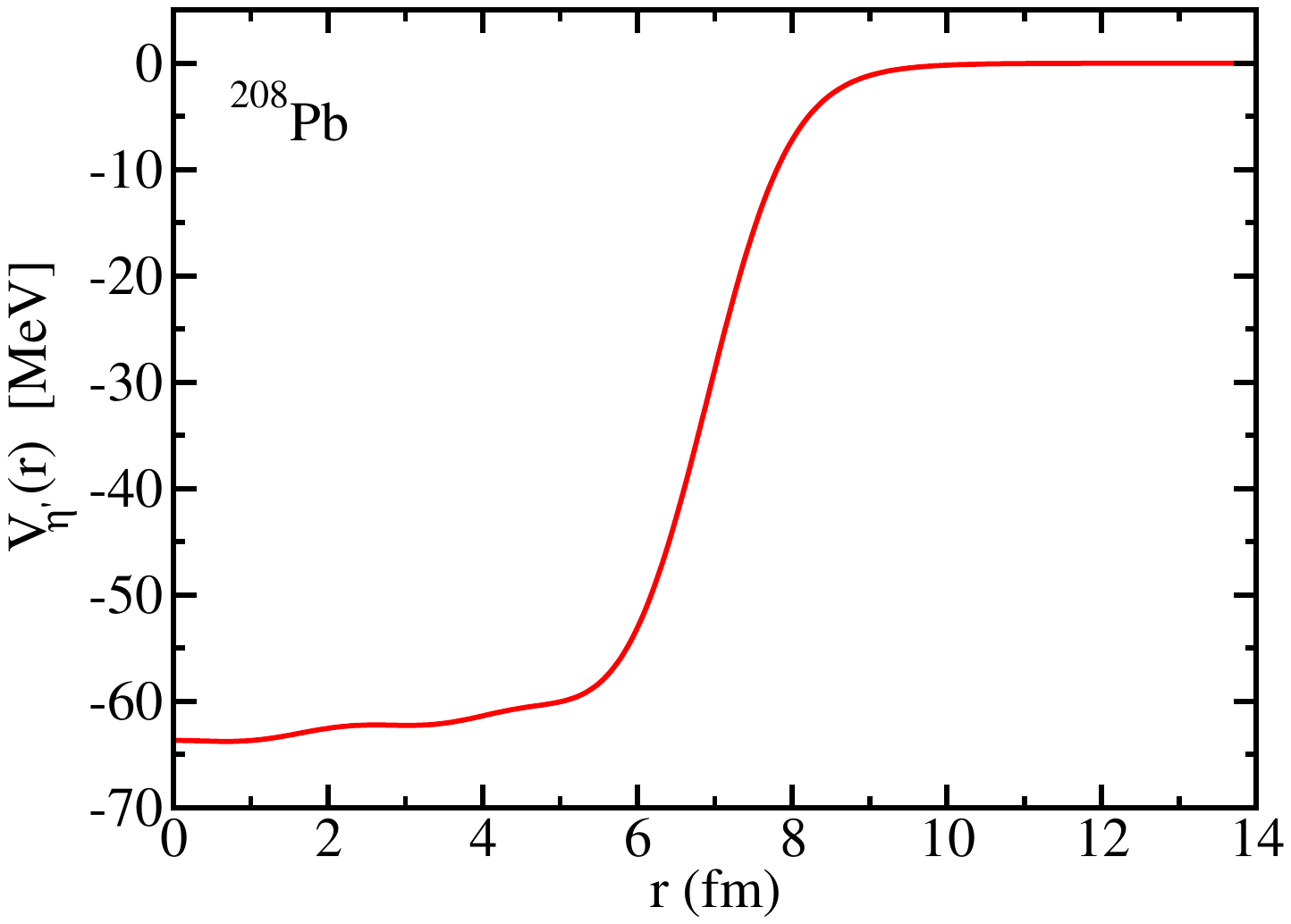} 
 \end{tabular}
 }
\caption{\label{fig:v_etaprime_nucl} $\eta'$ scalar potentials for several nuclei.}
\end{figure}
 To study the interactions of the $\eta$ and $\eta'$ with nuclei in more detail,
 we now consider the
 bound states of these mesons with nuclei when these mesons have been produced inside
 a nucleus $A$ and study the nuclear bound states, the so called mesic nuclei,  in a wide range of
 nuclear masses, namely for the nuclei listed in the previous section.

\section{\label{sec:nuclear_bound_states} Numerical results for the {\bm$\eta$}--
and {\bm$\eta'$}--nucleus bound  state energies}
  
In order to obtain the $\eta$ and $\eta'$ single-particle
energies in nuclei, and also to have
an idea of the relativistic effects, we solve numerically the Schr\"{o}dinger
and Klein-Gordon equations for these mesons with the scalar nuclear potential
given in \eqn{eqn:Vs} for various nuclei.
 
 We first solve the Schr\"{o}dinger equation (SE),
\begin{equation}
\label{eqn:sch}
\left(-\frac{1}{2 m}\nabla^{2} + V(\vec{r})\right)\psi(\vec{r})
= E\psi(\vec{r}),
\end{equation}
\noindent  where $V(\vec{r}) = V(r)$ is the scalar (nuclear) potential, given by \eqn{eqn:Vs},
$r = |\vec{r}|$ is the distance from the center of the nucleus,
and  $m$ is the reduced mass of the
$h$-nucleus system in vacuum, with $h=\eta, \eta'$.

Next we solve the Klein-Gordon equation (KGE,
\begin{equation}
\label{eqn:kg}
\left(-\nabla^{2} + (m + V(\vec{r}))^2\right)\phi(\vec{r})
= \mathcal{E}^2\psi(\vec{r}),
\end{equation}
where, as before, $V(\vec{r}) = V(r)$ is the scalar (nuclear) potential,
given by \eqn{eqn:Vs}, $r = |\vec{r}|$ is the distance from
the center of the nucleus, and $m$ is the reduced mass of the
$h$-nucleus system in vacuum, with $h=\eta, \eta'$.
In this case, the bound state energies ($E$) of
the $h$-nucleus system are given by $E = \mathcal{E}-m$,
where $\mathcal{E}$ is the energy eigenvalue in \eqn{eqn:kg}.

For the solution of Eqs.~(\ref{eqn:sch}) and (\ref{eqn:kg}), we use momentum space methods.
Here, these equations are first converted to a momentum space representation via a  Fourier 
transform, followed by a partial wave decomposition of the Fourier-transformed potential.  
Then, for a given value of angular momentum ($\ell$), the eigenvalues of the resulting  equation 
are found by the inverse iteration eigenvalue algorithm.
We note that at this point there is no advantage in using momentum space methods.  
However, the advantage will be apparent later, when we add an imaginary part to the nuclear
potential, in order to simulate absorption of the $\eta$ and $\eta'$ mesons by nuclei.

The results obtained for the single-particle energies $E$ for the $\eta$ and $\eta'$ are listed
 in Tables~\ref{tab:eta-A-bse} and \ref{tab:etaprime-A-bse}, respectively, for all nuclei listed
 in the previous section.
In \tab{tab:eta-A-bse}, we show the $\eta$-nucleus bound state energies obtained by
solving the Schr\"{o}dinger and Klein-Gordon equations. For each nucleus we have
computed all bound states but have listed only a few of them (up to four).
However, we note that the number of bound states increases with the mass of the nucleus, such
 that for the heavier nuclei we have a richer structure of bound states. Furthermore, we note
 that the relativistic corrections decrease the bound state energies for the $\eta$ by
 approximately 2 MeV. 
 
In \tab{tab:etaprime-A-bse}, we show the $\eta'$-nucleus bound state energies obtained by
solving the Schr\"{o}dinger  and Klein-Gordon equations.
As in the $\eta$ case, for each nucleus we have listed up to four bound states. 
We see that, compared to the $\eta$ meson, for the $\eta'$ we have a richer structure of 
bound states due to the fact that the $\eta'$ is heavier.
Finally, we note that the relativistic corrections are smaller for the $\eta'$ meson, by
 approximately  1 MeV, due to its larger mass. 
Thus, from Tables~\ref{tab:eta-A-bse} and \ref{tab:etaprime-A-bse}, we conclude that the
$\eta$ and $\eta'$ are expected to form mesic nuclei with all the nuclei considered. 
However, this is not a definite conclusion since, so far, we have not taken into account 
the abosorption of these mesons by nuclei.

\begin{table}[ht]
\begin{center}
\scalebox{0.9}{
\begin{tabular}{ll|rr}
  \hline \hline
& $n\ell$ & $E$ (SE) & $E$ (KGE)\\ 
  \hline
  $^{4}_{\eta}\text{He}$
  & $1s$ & -13.24 &  -10.99 \\  
  \hline
  $^{12}_{\eta}\text{C}$
  & $1s$ & -27.03 &  -25.25 \\
  &  $1p$ & -1.75 &  -0.87 \\ 
  \hline  
$^{16}_{\eta}\text{O}$
  & $1s$ & -32.60 &  -30.78 \\
  & $1p$ & -8.04 &  -6.47 \\
  \hline
 $^{40}_{\eta}\text{Ca}$
  & $1s$ & -48.71 &  -46.93 \\
  & $1p$ & -29.31 & -26.93 \\
  & $1d$ & -8.73 &  -6.67 \\
   & $2s$ & -7.16 & -5.43 \\     
  \hline
$^{48}_{\eta}\text{Ca}$
  & $1s$ & -49.34 &  -47.78 \\
  & $1p$ & -32.22 &  -29.97 \\
  & $1d$ & -13.31 & -11.08 \\
   & $2s$ & -10.34 & -8.7 \\ 
  \hline
$^{90}_{\eta}\text{Zr}$
  & $1s$ & -53.71 & -52.56 \\
  & $1p$ & -41.76 &  -39.85 \\
  & $1d$ & -27.71 & -25.32\\
  & $2s$ & -23.43 &  -21.04 \\
  \hline
$^{197}_{\eta}\text{Au}$ 
& $1s$  & -55.85 &  -55.12 \\
& $1p$  & -48.45 &  -47.13 \\
& $1d$ & -39.48 &  -37.60 \\
& $2s$  & -36.04 &  -34.01 \\
  \hline
$^{208}_{\eta}\text{Pb}$
& $1s$ & -57.59 & -56.85 \\
& $1p$ & -50.27 &  -48.92 \\
& $1d$ & -41.40 &  -39.81 \\
& $2s$ & -38.03 &  -35.95 \\
 \hline \hline
\end{tabular}
}
 \caption{\label{tab:eta-A-bse} Bound state energies ($E$) of the $\eta$ meson in
  nucleus of mass number $A$ obtained by solving the Schr\"odinger and Klein-Gordon
  equations. $E$ is in MeV.}
\end{center}
\end{table}

\begin{table}[ht]
\begin{center}
\scalebox{0.90}{
\begin{tabular}{ll|rr}
  \hline \hline
& $n\ell$ & $E$ (SE) & $E$ (KGE) \\
  \hline
  $^{4}_{\eta'}\text{He}$
  &  $1s$ & -23.72 &  -22.11 \\
    \hline
  $^{12}_{\eta'}\text{C}$
  &  $1s$ & -34.79 &  -33.88 \\
  & $1p$ & -13.63 & -12.72 \\
     \hline
  $^{16}_{\eta'}\text{0}$
  & $1s$ & -39.47 & -38.64 \\
   & $1p$ & -20.81 & -19.75 \\
  & $2s$ & -1.88 & -1.39 \\
  & $1d$ & -1.76 & -0.33 \\
  \hline
  $^{40}_{\eta'}\text{Ca}$
  & $1s$ & -53.09 & -52.38 \\
  & $1p$ & -39.52 &  -38.41 \\
  & $1d$ &-24.41 &  -23.12 \\
  & $2s$ & -21.66 &  -20.38 \\
  \hline  
  $^{48}_{\eta'}\text{Ca}$
  & $1s$ & -53.01 & -52.40 \\
   & $1p$ & -41.29 &  -40.30 \\
    & $1d$ & -27.90 &  -26.68 \\
  & $2s$ & -24.69 &  -23.45 \\
  \hline
 $^{90}_{\eta'}\text{Zr}$
& $1s$ & -55.62 &  -55.20 \\
& $1p$ & -47.79 &   -47.05 \\
& $1d$ & -38.47 &  -37.42 \\
& $2s$ & -35.31 &   -34.19 \\
\hline
$^{197}_{\eta'}\text{Au}$
& $1s$ & -56.28 &  -56.03 \\
& $1p$ & -51.59 &  -51.12\\
& $1d$ & -45.87 &  -45.15 \\
& $2s$ & -43.60 &  -42.80 \\
\hline
 $^{208}_{\eta'}\text{Pb}$
  & $1s$ & -57.90 &  -57.65 \\
  & $1p$ & -53.26 & -52.77 \\
   & $1d$ & -47.60 &  -46.87 \\
 & $2s$ & -45.38 &  -44.56 \\
 \hline \hline
\end{tabular}
}
  \caption{\label{tab:etaprime-A-bse}  Bound state energies ($E$) of the $\eta'$ 
  meson in nucleus of mass number $A$ obtained by solving the Schr\"odinger and
   Klein-Gordon equations. $E$ is in MeV.}
\end{center}
\end{table}

\section{\label{sec:nuclear_bound_states_complex}
Single-particle energies and absorption widths for the {\bm$\eta$} and {\bm$\eta'$} in nuclei.}

In order to make our results more realistic, we now add an important feature of the 
meson-nucleus interaction that was neglected in Ref.~\cite{Tsushima:1998qp} for the $\eta'$,
namely meson absorption in nuclei, which requires a complex potential.
We follow Refs.}~\cite{Tsushima:1998qw,Tsushima:1998qp} and construct this potential
phenomenologically by adding an imaginary part $W_{hA}(r)$ to the  $\eta$ and $\eta'$
scalar potentials discussed in the previous section, given in \eqn{eqn:Vs}, as follows
\begin{equation}
\label{eqn:Vs_complex}
V_{h A}(r)= \Delta m_{h}(\rho^{A}_{B}(r)) + i W_{hA}(r),
\end{equation}
where $W_{hA}(r)$ is related to  the absoprtion of the meson $h$ in the nuclear medium,
and is given by
\begin{eqnarray}
\label{eqn:imag_Vs}
W_{hA}(r)&=& -\dfrac{1}{2}\Gamma_{hA}(r),  \\
\Gamma_{hA}(r)&=&-\gamma\Delta m_{h}(\rho^{A}_{B}(r)) + \Gamma_{h}^{\text{vac}}.
\end{eqnarray}
Here $\Gamma_{h}^{\text{vac}}$ the meson decay width in vacuum
($\Gamma_{\eta}^{\text{vac}}=1.31$ keV and $\Gamma_{\eta'}^{\text{vac}}=0.188$ 
MeV~\cite{Workman:2022ynf}), and $\gamma$ is a phenomenological parameter used to 
simulate the strength of the absoprtion of the meson in the nuclear medium. 

Next, we calculate the bound state energies and widths for the $\eta$- and $\eta'$-mesic nuclei
using the Schr\"odinger and Klein-Gordon equations with the complex potential, \eqn{eqn:Vs_complex}, for several values of the parameter
$\gamma$,  which cover the estimated  widths of the $\eta$ and $\eta'$ mesons in the nuclear
medium quoted in the introduction section.

For the moment we ignore $\Gamma_{h}^{\text{vac}}$. However, since it is very small it will not
contribute much.
In Tables \ref{tab:eta-A-se-complex}-\ref{tab:eta-A-kge-complex}, we show the
results for tbe  bound state energies ($E$) and full widths ($\Gamma$) of the $\eta$-mesic nuclei
of mass number $A$, obtained by solving the Schr\"odinger and  Klein-Gordon equations,
respectively, for various values of the strenght of the imaginary part of the potential
$\gamma= 0.0,\, 0.25,\,0.5,\, 1.0$. 
The bound state energies and full widths are obtained from the complex energy eigenvalue 
$\mathcal{E}$ as $\mathcal{E}= E - i\Gamma/2$ for the Schr\"odinger equation and as 
$\mathcal{E}= E  + m - i\Gamma/2$ for the Klein-Gordon equation. 
In Tables \ref{tab:etaprime-A-se-complex}-\ref{tab:etaprime-A-kge-complex} we
present the corresponding results for the $\eta'$ meson.
Note that the results for $\gamma=0$, for both the $\eta$ and $\eta'$ mesons, correspond to
 the results given in Tables~\ref{tab:eta-A-bse} and \ref{tab:etaprime-A-bse}, which we have 
 already  discussed in the previous section.
 
We now discuss the results obtained when the imaginary part of the potential is included,
columns with $\gamma= 0.25,\,0.5$ and $\gamma= 1.0$ in Tables
\ref{tab:eta-A-se-complex}-\ref{tab:etaprime-A-kge-complex}.
The following conclusions, obtained from Tables 
\ref{tab:eta-A-se-complex}-\ref{tab:etaprime-A-kge-complex}, apply both to the $\eta$- and $\eta'$-mesic nuclei.
For now, we will only discuss the ground states since all mesic nuclei have at least one bound state.

Adding an absorptive part of the potential changes the situation appreciably. 
The effects are larger the larger $\gamma$ is for both the non-relativistc and relativistic cases. 
Clearly, the imaginary part of the potential is repulsive, being more repulsive for $\gamma=1$.
Whether or not  the bound states can be observed experimentally is sensitive to the value
of the parameter $\gamma$ since $\Gamma$ increases with increasing $\gamma$.
Furthermore, since the so-called dispersive effect of the absorptive potential is repulsive, the 
binding energies for all nuclei decrease with $\gamma$. However, they decrease very little. 
Even for the largest value of $\gamma$, there is at least one bound state. We have found similar
results for the $\phi$ meson in a previous paper~\cite{Cobos-Martinez:2017woo}. 
Note that the width of the ground state increases with $\gamma$ for all nuclei, as expected, since 
a larger $\gamma$ means that the strength of the imaginary part of the potential is larger. 

$\eta$ and $\eta'$-mesic bound states have been studied by several authors
in a variety of approaches, such as the quark-meson coupling (QMC)
model~\cite{Bass:2005hn, Bass:2013nya}, chiral coupled channels~\cite{Nagahiro:2011fi},
the Nambu--Jona-Lasinio model~\cite{Bernard:1987sx,Nagahiro:2006dr,Costa:2002gk},
and linear $\sigma$ model~\cite{Sakai:2013nba,Sakai:2022xao}, and chiral unitary approach
using a microscopic many-body theory~\cite{Inoue:2002xw,Garcia-Recio:2002xog},
as well as other more formal approaches~\cite{Nagahiro:2004qz,Jido:2011pq}.
It is interesting to compare our results with those of
Refs.~\cite{Inoue:2002xw,Garcia-Recio:2002xog}, where both real and imaginary parts of the potential
have been evaluated from the $\eta$ self-energy in nuclei in chiral unitary approach.
For the nuclei for which the comparison is possible ($^{12}_{\eta}\text{C}$,
$^{40}_{\eta}\text{Ca}$, and $^{208}_{\eta}\text{Pb}$),
we see that our bound state energies are more than twice theirs
(see Table 1 Ref.~\cite{Garcia-Recio:2002xog}) when we compare with their results
obtained using an energy-dependent potential; however, the situation improves to our
favor when we compare with their results obtained using an energy-independent potential.
The fact that our results compare better with those of Ref.~\cite{Garcia-Recio:2002xog}
seems to indicate that we can improve our results, and reduce the amount of binding energy,
by taking into account the energy and momentum dependence of the $\eta$ (and $\eta'$)
self-energy.
The main goal of the present work is to improve the previous results
by adding  an imaginary part to the potential in order to assess the feasibility of observation
of the $\eta$- and $\eta'$-mesic nuclei predicted by our approach.
In any case, it is possible that our results give overbinding. This may be based on the results
from Ref.~\cite{Xie:2018aeg} where no clear  $^{4}_{\eta}\text{He}$ bound states are found.
The analysis in Ref.~\cite{Xie:2018aeg} is carried by analyzing the data on the cross section
for the $dd\to \eta^4\text{He}$ reaction close to threshold.
However, the present approach is based on the mean field approach, and then
the results for lighter nuclei may not be appropriate for $^{4}_{\eta}\text{He}$.
In fact, the nucleon density distribution for $^{4}\text{He}$ is not calculated here, which 
is the basis to calculate the $^{4}_{\eta}\text{He}$ potential. 
Furthermore, in the case when  the width of the $\eta$ (imaginary part of the potential) is indeed large in reality, the conclusion drawn based on the experimental data  that they found no bound 
$\eta$-nucleus states,  may be changed. That is why we have addressed the calculation with a wider 
range of the imaginary part of the potential.
 
\begin{table}[t]
\begin{center}
\scalebox{0.85}{
\begin{tabular}{ll|rr|rr|rr|rr}
  \hline \hline
  & &  \multicolumn{2}{c}{$\gamma=0$}   & 
  \multicolumn{2}{c}{$\gamma=0.25$} &
  \multicolumn{2}{c}{$\gamma=0.5$} & 
  \multicolumn{2}{c}{$\gamma=1.0$}  \\
  \hline
& $n\ell$ & $E$ & $\Gamma$ & $E$ & $\Gamma$ & $E$ & $\Gamma$ 
& $E$ & $\Gamma$  \\
  \hline
  $^{4}_{\eta}\text{He}$
  & $1s$ & -13.24 & 0 & -12.97 & 9.81 &  -12.21 & 19.86 & -9.45 & 41.45 \\  
  \hline
  $^{12}_{\eta}\text{C}$
  & $1s$ & -27.03 & 0 & -26.92 &  11.77 &  -26.60 & 23.64 & -25.44 & 47.97 \\
  &  $1p$ & -1.75 & 0 &  -1.33 &  6.05 & N & N & N &  N \\ 
  \hline  
$^{16}_{\eta}\text{O}$
  & $1s$ & -32.60 & 0 & -32.51 & 12.84 &   -32.27 & 25.77  & -31.39 & 52.09  \\
  & $1p$ & -8.04 & 0 &  -7.80 &  8.99 &  -7.12 & 18.30 &  -4.84 & 38.46 \\
  \hline
 $^{40}_{\eta}\text{Ca}$
  & $1s$ & -48.71 &  0 &  -48.66 & 15.86 & -48.52 & 31.76 & -48.00 &63.84 \\
  & $1p$ & -29.31 & 0 &  -29.20 & 13.83 & -28.88 &  27.78  & -27.75 & 56.30\\
  & $1d$ & -8.73 &  0 &  -8.50 &  10.92 & -7.84 & 22.14 & -5.64 & 46.09\\
   & $2s$ & -7.16 & 0 &  -6.79 & 9.01 & -5.77 & 18.56 & N & N \\     
  \hline
$^{48}_{\eta}\text{Ca}$
  & $1s$ & -49.34 & 0 &  -49.31 & 15.61  &-49.19 & 31.25  & -48.78 & 62.77 \\
  & $1p$ & -32.22 &  0 &  -32.13 & 14.03  &-31.88 & 28.15 & -30.97 & 56.91\\
  & $1d$ & -13.31 & 0 &   -13.13 &  11.80 & -12.63 & 23.81 & -10.90 & 48.98\\
   & $2s$ & -10.34 & 0 &  -10.05 & 10.25 & -9.26 & 20.91 & N & N\\ 
  \hline
$^{90}_{\eta}\text{Zr}$
  & $1s$ & -53.71 & 0 &  -53.69 & 15.77 & -53.63 & 31.56 & -53.40 & 63.27 \\
  & $1p$ & -41.76 &  0 & -41.71 & 14.95 & -41.57 & 29.94  & -41.08 & 60.22 \\
  & $1d$ & -27.71 & 0 &  -27.62 & 13.85 & -27.36 & 27.79 & -26.49 & 56.21 \\
  & $2s$ & -23.43 &  0 &  -22.28 & 9.68 & -22.95 & 26.47  & -21.75 & 53.93 \\
  \hline
$^{197}_{\eta}\text{Au}$ 
& $1s$  & -55.85 & 0 &  -55.84 & 15.45 & -55.81 & 30.92  & -55.69 &61.91 \\
& $1p$  & -48.45 & 0 &  -48.43 & 15.07 & -48.36 & 30.16 & -48.13 & 60.47 \\
& $1d$ & -39.48 & 0 &  -39.44 & 14.56 & -39.32 & 29.17 & -38.91 & 58.62 \\
& $2s$  & -36.04 &  0 &  -35.99 & 14.29  & -35.84 & 28.63  & -35.31 & 57.66 \\
  \hline
$^{208}_{\eta}\text{Pb}$
& $1s$ & -57.59 & 0 &  -57.58 &  15.87 & -57.55 & 31.76 & -57.44 & 63.58\\
& $1p$ & -50.27 & 0 &  -50.25 & 15.49 & -50.18 & 30.99  & -49.95 & 62.14  \\
& $1d$ & -41.40 & 0 &  -41.36 & 14.99 & -41.25 & 30.01 & -40.85 & 60.30 \\
& $2s$ & -38.03 & 0 &  -37.98 & 14.72 & -37.83 &29.50 & -37.32 & 59.37 \\
 \hline \hline
\end{tabular}
}
  \caption{\label{tab:eta-A-se-complex} Bound state energies ($E$) 
  and full widths ($\Gamma$) of $\eta$ meson in nucleus of mass 
  number $A$ obtained by solving the Schr\"odinger for various values of $\gamma$. 
}
\end{center}
\end{table}

\begin{table}[t]
\begin{center}
\scalebox{0.85}{
\begin{tabular}{ll|rr|rr|rr|rr}
  \hline \hline
  & &  \multicolumn{2}{c}{$\gamma=0$}   &
  \multicolumn{2}{c}{$\gamma=0.25$}   &
  \multicolumn{2}{c}{$\gamma=0.5$} & \multicolumn{2}{c}{$\gamma=1.0$}  \\
  \hline
& $n\ell$ & $E$ & $\Gamma$ & $E$ & $\Gamma$ & $E$ & $\Gamma$ & $E$ & 
$\Gamma$ \\
  \hline
  $^{4}_{\eta}\text{He}$
 & $1s$ &  -10.99 & 0 & -10.79 &  8.21 & -10.20 & 16.65 & -8.13 & 34.94 \\
  \hline
$^{12}_{\eta}\text{C}$
& $1s$ &  -25.25 & 0 & -25.16 & 10.86 & -24.91 & 21.82 & -24.02 &  44.29 \\
& $1p$ &   -0.87 & 0 & -0.43 & 4.97  & N & N & N & N \\
  \hline
$^{16}_{\eta}\text{O}$
& $1s$ &  -30.78 & 0 & -30.72 & 12.00 & -30.53 & 24.07 & -29.86 & 48.67 \\
& $1p$ &   -6.47 & 0 & -6.26 & 7.84 & -5.67 & 15.99  & -3.77 & 33.80 \\
  \hline
$^{40}_{\eta}\text{Ca}$
 & $1s$ &  -46.93 & 0 & -46.89 & 15.12 & -46.79 & 30.28 & -46.43 & 60.87 \\
& $1p$ &  -26.93 & 0 & -26.85 & 12.67 & -26.61 & 25.44 & -25.77 & 51.59 \\
& $1d$ &   -6.67 & 0 & -6.47 & 9.48 & -5.91 & 19.27  & -4.15 & 40.31 \\
& $2s$ &   -5.43 & 0 & -5.09 & 7.51 & -4.18 & 15.59 & N & N \\
  \hline
$^{48}_{\eta}\text{Ca}$
& $1s$ &  -47.78 & 0 & -47.75 & 14.98 & -47.66 & 30.00  & -47.38 & 60.25 \\
& $1p$ &  -29.97 & 0 & -29.90 & 12.99 & -29.71 & 26.06 & -29.04 & 52.71 \\
& $1d$ &  -11.08 & 0 &  -10.93 & 10.45  & -10.51 & 21.10 & -9.15 & 43.52 \\
& $2s$ &   -8.7 & 0 &  -8.11 & 8.83   & -7.42 & 18.06  & N & N \\
  \hline
  $^{90}_{\eta}\text{Zr}$
& $1s$ &  -52.56 & 0 & -52.54 & 15.34 & -52.50 & 30.71 & -52.34 & 61.56 \\
& $1p$ &  -39.85 & 0 & -39.81 & 14.17 & -39.71 & 28.40 & -39.36 & 57.11 \\
& $1d$ &  -25.32 & 0 & -25.25 & 12.74 & -25.06 & 25.57 & -24.40 & 51.75 \\
& $2s$ &  -21.04 & 0 & -20.94 & 11.95 & -20.65 & 24.04 & -19.70 & 49.03 \\
  \hline
$^{197}_{\eta}\text{Au}$
& $1s$ &  -55.12 & 0 & -55.11 & 15.20 & -55.09 & 30.41  & -55.01 & 60.89 \\
& $1p$ &  -47.13 & 0 & -47.11 & 14.58 & -47.06 & 29.19 & -46.90 & 58.53 \\
& $1d$ &  -37.60 & 0 & -37.58 & 13.83 & -37.49 & 27.69 & -37.20 & 55.67 \\
& $2s$ &  -34.01 & 0 & -33.97 & 13.45  & -33.86 & 26.96 & -33.47 & 54.31 \\
  \hline
$^{208}_{\eta}\text{Pb}$
& $1s$ &  -56.85 & 0 & -56.84 & 15.61 & -56.82 & 31.24 & -56.75 & 62.55 \\
& $1p$ &  -48.92 & 0 & -48.90 & 14.99 & -48.86 & 30.01  &  -48.70 &  60.17  \\
& $1d$ &  -39.81 & 0 & -39.45 &  14.24 & -39.37 & 28.51 & -39.09 & 57.29 \\
& $2s$ &  -35.95 & 0 & -35.91 & 13.87 & -35.80 & 27.80  & -35.43 & 55.96  \\
 \hline \hline
\end{tabular}
}
  \caption{\label{tab:eta-A-kge-complex} Bound state energies ($E$) 
  and full widths ($\Gamma$) of $\eta$ meson in nucleus of mass 
  number $A$ obtained by solving the Klein-Gordon equation for various values of 
  $\gamma$.
}
\end{center}
\end{table}

\begin{table}[t]
\begin{center}
\scalebox{0.85}{
\begin{tabular}{ll|rr|rr|rr|rr}
  \hline \hline
  & & \multicolumn{2}{c}{$\gamma=0$} & \multicolumn{2}{c}{$\gamma=0.25$} 
  & \multicolumn{2}{c}{$\gamma=0.5$}  & \multicolumn{2}{c}{$\gamma=1.0$} \\
  \hline
& $n\ell$ & $E$ & $\Gamma$ & $E$ & $\Gamma$  & $E$ & $\Gamma$ \\
  \hline
  $^{4}_{\eta'}\text{He}$
  &  $1s$ & -23.72 &   0 & -23.55 & 12.27  & -23.06 & 24.70 & -21.29 & 50.50 \\
    \hline
  $^{12}_{\eta'}\text{C}$
  &  $1s$ & -34.79 & 0 & -34.72 & 12.70  & -34.52 & 25.47 & -33.78 & 51.37 \\
  & $1p$ & -13.63 & 0 & -13.47 & 9.62  & -13.00 & 19.41 & -11.35 & 39.95 \\
     \hline
  $^{16}_{\eta'}\text{0}$
  & $1s$ & -39.47 & 0 & -39.42 & 13.40 & -39.27 & 26.86 & -38.74 & 54.04 \\
   & $1p$ & -20.81 & 0 & -20.69 & 11.31 & -20.35 & 22.73  & -19.16 & 46.28 \\
  & $2s$ & -1.88 & 0 & -1.33 & 5.22 & N & N & N & N \\
  & $1d$ & -1.76 & 0 & -1.43 & 7.89 & -0.35 & 13.61 & N & N \\
  \hline
  $^{40}_{\eta'}\text{Ca}$
  & $1s$ & -53.09 & 0 & -53.06 & 15.87 & -52.98 & 31.76 & -52.66 & 63.71 \\
  & $1p$ & -39.52 & 0 & -39.45 & 14.65 & -39.27 & 29.35 & -38.62 & 59.11 \\
  & $1d$ &-24.41 & 0 & -24.30 & 13.10 & -23.98 & 26.31 & -22.85 & 53.40 \\
  & $2s$ & -21.66 &  0 & -21.51 & 12.40 & -21.09 & 24.96 & -19.64 & 51.01\\
  \hline  
  $^{48}_{\eta'}\text{Ca}$
  & $1s$ & -53.01 & 0 &  -52.99 & 15.5 & -52.92 & 31.05 & -52.67 & 62.25 \\
   & $1p$ & -41.29 & 0 & -41.24 & 14.58 & -41.10 & 29.20 & -40.58 & 58.73 \\
     & $1d$ & -27.90 &  0 & -27.81 & 13.38 & -27.56 & 26.85 & -26.67 & 54.32 \\
  & $2s$ & -24.69 &  0 & -24.57 & 12.81 & -24.24 & 25.74 & -23.09 & 52.33 \\
  \hline
 $^{90}_{\eta'}\text{Zr}$
& $1s$ & -55.62 &   0 & -55.61 & 15.45 & -55.57 & 30.92 & -55.44 & 61.92 \\
& $1p$ & -47.79 &  0 & -47.77 & 14.98 & -47.69 & 29.98 & -47.42 & 60.13  \\
& $1d$ & -38.47 &  0 & -38.42 & 14.37 & -38.29 & 28.78 & -37.82 & 57.88 \\
& $2s$ & -35.31 &  0 & -35.25 & 14.07 & -35.08 & 28.21 & -34.49 & 56.84 \\
\hline
$^{197}_{\eta'}\text{Au}$
& $1s$ & -56.28 &  0 & -56.28 & 15.03 & -56.26 & 30.06 & -56.20 & 60.15 \\
& $1p$ & -51.59 &  0 &  -51.58 & 14.80 & -51.54 & 29.61 &  -51.41 & 59.31 \\
& $1d$ & -45.87 &  0 & -45.85 & 14.52 & -45.79 & 29.06 & -45.57 & 58.26  \\
& $2s$ & -43.60 &  0 & -43.57 & 14.39  & -43.50 & 28.80 & -43.02 & 58.09 \\
\hline
 $^{208}_{\eta'}\text{Pb}$
  & $1s$ & -57.90 & 0 & -57.90 & 15.42  & -57.88 & 30.84 & -57.81 & 61.73 \\
  & $1p$ & -53.26 & 0 & -53.24 & 15.19 & -53.21 & 30.40 & -53.08 & 60.88 \\
   & $1d$ & -47.60 & 0 & -47.57 & 14.91  & -47.51 & 29.84  & -47.29 & 59.83  \\
 & $2s$ & -45.38 &  0 & -45.35 & 14.78  & -45.27 & 29.59  & -45.01 & 59.36 \\
 \hline \hline
\end{tabular}
}
  \caption{\label{tab:etaprime-A-se-complex}  Bound state energies ($E$) 
  and full widths ($\Gamma$) of $\eta'$ meson in nucleus of mass 
  number $A$ obtained by solving the Schr\"{o}dinger equation
  for various values of $\gamma$.
}
\end{center}
\end{table}

\begin{table}[t]
\begin{center}
\scalebox{0.85}{
\begin{tabular}{ll|rr|rr|rr|rr}
  \hline \hline
  & &  \multicolumn{2}{c}{$\gamma=0$}   &  \multicolumn{2}{c}{$\gamma=0.25$} &
  \multicolumn{2}{c}{$\gamma=0.5$} & \multicolumn{2}{c}{$\gamma=1.0$}  \\
  \hline
  & $n\ell$ & $E$ & $\Gamma$ & $E$ & $\Gamma$ & $E$ & $\Gamma$ &
  $E$ & $\Gamma$ \\
  \hline
  $^{4}_{\eta'}\text{He}$
 & $1s$ &  -22.11 & 0 & -21.96 & 11.37 & -21.55 & 22.89  & -20.06 &  46.83  \\
  \hline
$^{12}_{\eta'}\text{C}$
& $1s$ &  -33.88 & 0 & -33.82 & 12.30 &  -33.64 & 24.66 & -33.00 & 49.73  \\
& $1p$ &  -12.72 & 0 & -12.57 & 9.06 &  -12.15 & 18.29  & -10.67 & 37.68  \\
  \hline
$^{16}_{\eta'}\text{O}$
& $1s$ &  -38.64 & 0 & -38.59 & 13.06 &  -38.46 & 26.17 & -38.00 & 52.65  \\
& $1p$ &  -19.75 & 0 & -19.65 & 10.76 &  -19.34 & 21.64  & -18.28 & 44.07 \\
& $2s$ &   -1.39 & 0 & -0.84 & 4.48 &  N & N & N & N \\
& $1d$ &   -0.33 & 0 & -0.69 & 7.20 &  N & N & N & N \\
  \hline
$^{40}_{\eta'}\text{Ca}$
 & $1s$ &  -52.38 & 0 & -52.35 & 15.59 &  -52.28 & 31.22 & -52.00 & 62.61 \\
 & $1p$ &  -38.41 & 0 & -38.35 & 14.18 &  -38.19 & 28.41 & -37.63 & 57.22 \\
 & $1d$ &  -23.12 & 0 & -23.02 & 12.46 &  -22.74 & 25.03 & -21.75 & 50.81  \\
& $2s$ &  -20.38 & 0 & -20.25 & 11.72 &  -19.87 & 23.60 & -18.58 & 48.25  \\
  \hline
$^{48}_{\eta'}\text{Ca}$
& $1s$ &  -52.40 & 0 & -52.38 & 15.29 &  -52.32 & 30.60  & -52.11 & 61.35  \\
& $1p$ &  -40.30 & 0 & -40.26 & 14.18 &  -40.13 & 28.40 & -39.68 & 57.12  \\
& $1d$ &  -26.68 & 0 & -26.59 & 12.82 &  -26.37 & 25.72 & -25.58 & 52.02  \\
& $2s$ &  -23.45 & 0 & -23.34 & 12.19 &  -23.04 & 24.51  & -22.01 & 49.85  \\
  \hline
  $^{90}_{\eta'}\text{Zr}$ 
& $1s$ &  -55.20 & 0 & -55.19 & 15.31 &  -55.16 & 30.63  & -55.04 & 61.35  \\
& $1p$ &  -47.05 & 0 & -47.02 & 14.70  &  -46.96 & 29.43  & -46.72 & 59.04  \\
& $1d$ &  -37.42 & 0 & -37.38 & 13.96  &  -37.27 & 27.96  & -36.86 & 56.22 \\
& $2s$ &  -34.19 & 0 & -34.14 & 13.61 &  -33.99 & 27.29  & -33.47 & 54.98  \\
  \hline
$^{197}_{\eta'}\text{Au}$
& $1s$ &  -56.03 & 0 & -56.03 & 14.94  &  -56.01 & 29.89  & -55.96 & 59.83  \\
& $1p$ &  -51.12 & 0 & -51.10 & 14.64 &  -51.07 & 29.30  & -50.96 & 58.67 \\
& $1d$ &  -45.15 & 0 & -45.14 & 14.27 &  -45.08 & 28.56  & -44.89 & 57.26  \\
& $2s$ &  -42.80 & 0 & -42.78 & 14.10  &  -42.71 & 28.22  & -42.47 & 56.63  \\
  \hline
$^{208}_{\eta'}\text{Pb}$
& $1s$ &  -57.65 & 0 & -57.64 & 15.34 &  -57.63 & 30.68  & -57.57& 61.40 \\
& $1p$ &  -52.77 & 0 & -52.76 & 15.03 &  -52.73 & 30.07  & -52.62 & 60.23  \\
& $1d$ &  -46.87 & 0 & -46.85 & 14.66 &  -46.80 & 29.33  & -46.61 & 58.80  \\
& $2s$ &  -44.56 & 0 & -44.54 & 14.49 &  -44.47 & 29.00  & -44.24 & 58.19 \\
 \hline \hline
\end{tabular}
}
 \caption{\label{tab:etaprime-A-kge-complex} Bound state energies ($E$) 
  and full widths ($\Gamma$) of $\eta'$ meson in nucleus of mass 
  number $A$ obtained by solving the Klein-Gordon equation for various values of 
  $\gamma$. 
}
\end{center}
\end{table}

\section{\label{sec:conclusions}Summary and conclusions}

We have updated the mass shift of the $\eta$ and $\eta'$ mesons in symmetric nuclear
matter, using the most up to date mixing angle $\theta_P= -11.3^{\circ}$,
within the quark-meson coupling model.
Using these mass shift as input, we have calculated, in the local
density approximation, the
real part of the scalar nuclear potential for  the $\eta$ and $\eta'$ mesons for various nuclei,
covering a wide range of nuclear masses.
The nuclear density distributions for all nuclei, except the lightest one, were computed using the
quark-meson coupling model self-consistently.
We found that these potentials are attractive in all cases.
Then we calculated the bound state energies for the $\eta$- and $\eta'$-mesic nuclei for
several nuclei, by solving the Schr\"odiger and Klein-Gordon equations.
We found that by neglecting meson absorption by nuclei, these mesons should form 
mesic nuclei  with all the nuclei considered. 
Even though this is an important step, it ignores the absorption of these mesons by
nuclei. To remedy this, we have added, in a phenomenological way,
an imaginary part to the $\eta$ and $\eta'$-nucleus potentials with wider ranges,
and again solved the Schr\"odiger and
Klein-Gordon equations with complex potentials. Our results show that the 
$\eta$ and $\eta'$ mesons are expected to form mesic nuclei with all the nuclei considered.
The main feature of forming bound states with nuclei is not changed by the introduction of imaginary part of the potentials, though the feasibility for experiment
is certainly influenced.
Namely, the signal for the formation of the $\eta$- and $\eta'$ mesic nuclei may be difficult
to identify experimentally, given the similarity between the real and imaginary parts (widths) 
of the bound state energies. Furthermore, our results depend on the strength of the imaginary
part of the meson-nucleus potential, as expected. Thus, in order to quantify
this uncertainty and the sensitivity of our results to its value,
we have analyzed three values of strength for the imaginary part of the meson-nucleus potential.
Therefore, the feasibility of observation of the $\eta$- and $\eta'$-mesic nuclei
needs further investigation.

\section*{Acknowledgements}
The authors acknowledge the support and warm hospitality of APCTP
(Asia Pacific Center for Theoretical Physics) during the Workshop (APCTP PROGRAMS 2023)
''Origin of Matter and Masses in the Universe: Hadrons in free space, dense nuclear medium,
and compact stars'', where part of the present work was presented and discussed.
The authors also thank the OMEG (Origin of Matter and Evolution of Galaxies) Institute
at Soongsil University for the supports in many aspects
during the collaboration visit in Korea, and ''70th OMEG-SSANP Workshop''.
JJCM acknowledges financial support fromthe University of Sonora under grant
USO315007861.
K.~T.~was supported by Conselho Nacional de Desenvolvimento
Cient\'{i}fico e Tecnol\'ogico (CNPq, Brazil), Processes No.~313063/2018-4
and No.~426150/2018-0, and FAPESP Process No.~2019/00763-0 and No. 2023/07313-6,
and his work was also part of the projects,
Instituto Nacional de Ci\^{e}ncia e
Tecnologia - Nuclear Physics and Applications
(INCT-FNA), Brazil, Process No.~464898/2014-5,
and FAPESP Tem\'{a}tico, Brazil, Process No.~2017/05660-0.


\begin{thebibliography}{100}


\bibitem{piAF:2022gvw}
T.~Nishi \textit{et al.} [piAF],
``Chiral symmetry restoration at high matter density observed in pionic atoms,''
Nature Phys. \textbf{19}, no.6, 788-793 (2023)
[arXiv:2204.05568 [nucl-ex]].

\bibitem{Krein:2017usp}
G.~Krein, A.~W.~Thomas and K.~Tsushima,
``Nuclear-bound quarkonia and heavy-flavor hadrons,''
Prog. Part. Nucl. Phys. \textbf{100}, 161-210 (2018)
[arXiv:1706.02688 [hep-ph]].

\bibitem{Metag:2017yuh}
V.~Metag, M.~Nanova and E.~Y.~Paryev,
``Meson\textendash{}nucleus potentials and the search for meson\textendash{}nucleus bound states,''
Prog. Part. Nucl. Phys. \textbf{97}, 199-260 (2017)
[arXiv:1706.09654 [nucl-ex]].

\bibitem{Khreptak:2023lbh}
A.~Khreptak, M.~Skurzok and P.~Moskal,
``Search for \ensuremath{\eta}-mesic nuclei: a review of experimental and theoretical advances,''
Front. in Phys. \textbf{11}, 1186457 (2023)

\bibitem{Bass:2021rch}
S.~D.~Bass, V.~Metag and P.~Moskal,
``The $\eta$- and $\eta'$-nucleus interactions and the search for $\eta$, $\eta'$- mesic states,''
[arXiv:2111.01388 [hep-ph]].

\bibitem{Haider:2015fea}
Q.~Haider and L.~C.~Liu,
``Eta-mesic nuclei: past, present, future,''
Int. J. Mod. Phys. E \textbf{24}, no.10, 1530009 (2015)
[arXiv:1509.05487 [nucl-th]].

\bibitem{Bass:2018xmz}
S.~D.~Bass and P.~Moskal,
``$\eta'$ and $\eta$ mesons with connection to anomalous glue,''
Rev. Mod. Phys. \textbf{91}, no.1, 015003 (2019)
[arXiv:1810.12290 [hep-ph]].

\bibitem{Kelkar:2013lwa}
N.~G.~Kelkar, K.~P.~Khemchandani, N.~J.~Upadhyay and B.~K.~Jain,
``Interaction of eta mesons with nuclei,''
Rept. Prog. Phys. \textbf{76}, 066301 (2013)
[arXiv:1306.2909 [nucl-th]].

\bibitem{Bass:2005hn}
S.~D.~Bass and A.~W.~Thomas,
``eta bound states in nuclei: A Probe of flavor-singlet dynamics,''
Phys. Lett. B \textbf{634}, 368-373 (2006)
[arXiv:hep-ph/0507024 [hep-ph]].

\bibitem{Li:2022dry}
J.~Li, J.~Gui and P.~Zhuang,
``symmetry restoration at high baryon density,''
Chin. Phys. C \textbf{47}, no.10, 104102 (2023)
[arXiv:2209.13190 [hep-ph]].

\bibitem{Haider:1986sa}
Q.~Haider and L.~C.~Liu,
``Formation of an $\eta$ Mesic Nucleus,''
Phys. Lett. B \textbf{172}, 257-260 (1986)

\bibitem{Skurzok:2018paa}
M.~Skurzok, P.~Moskal, N.~G.~Kelkar, S.~Hirenzaki, H.~Nagahiro and N.~Ikeno,
``Constraining the optical potential in the search for $\eta$-mesic $^4$He,''
Phys. Lett. B \textbf{782}, 6-12 (2018)
[arXiv:1802.08597 [nucl-ex]].

\bibitem{Ikeno:2017xyb}
N.~Ikeno, H.~Nagahiro, D.~Jido and S.~Hirenzaki,
``$\eta$-nucleus interaction from the $d+d$ reaction around the $\eta$ production threshold,''
Eur. Phys. J. A \textbf{53}, no.10, 194 (2017)
[arXiv:1708.07692 [nucl-th]].

\bibitem{CBELSATAPS:2016qdi}
M.~Nanova \textit{et al.} [CBELSA/TAPS],
``Determination of the real part of the $\eta$'-Nb optical potential,''
Phys. Rev. C \textbf{94}, no.2, 025205 (2016)
[arXiv:1607.07228 [nucl-ex]].

\bibitem{CBELSATAPS:2013waf}
M.~Nanova \textit{et al.} [CBELSA/TAPS],
``Determination of the ${\eta}$'-nucleus optical potential,''
Phys. Lett. B \textbf{727}, 417-423 (2013)
[arXiv:1311.0122 [nucl-ex]].

\bibitem{CBELSATAPS:2018sck}
M.~Nanova \textit{et al.} [CBELSA/TAPS],
``The $\eta^{\prime}$ -carbon potential at low meson momenta,''
Eur. Phys. J. A \textbf{54}, no.10, 182 (2018)
[arXiv:1810.01288 [nucl-ex]].

\bibitem{CBELSATAPS:2012few}
M.~Nanova \textit{et al.} [CBELSA/TAPS],
``Transparency ratio in \ensuremath{\gamma}A\textrightarrow{}\ensuremath{\eta'}A' and the in-medium \ensuremath{\eta'} width,''
Phys. Lett. B \textbf{710}, 600-606 (2012)
[arXiv:1204.2914 [nucl-ex]].

\bibitem{Friedrich:2016cms}
S.~Friedrich, M.~Nanova, V.~Metag, F.~N.~Afzal, D.~Bayadilov, B.~Bantes, R.~Beck, M.~Becker, S.~B\"ose and K.~T.~Brinkmann, \textit{et al.}
``Momentum dependence of the imaginary part of the $ \omega$ - and $ \eta^{\prime}$ -nucleus optical potential,''
Eur. Phys. J. A \textbf{52}, no.9, 297 (2016)
[arXiv:1608.06074 [nucl-ex]].

\bibitem{LEPS2BGOegg:2020cth}
N.~Tomida \textit{et al.} [LEPS2/BGOegg],
``Search for $\eta'$ bound nuclei in the $^{12}{\rm C}(\gamma,p)$ reaction with simultaneous detection of decay products,''
Phys. Rev. Lett. \textbf{124}, no.20, 202501 (2020)
[arXiv:2005.03449 [nucl-ex]].
 
\bibitem{Fujioka:2020ewc}
H.~Fujioka, K.~Itahashi, V.~Metag, M.~Nanova and Y.~K.~Tanaka,
``Comment on \textquotedblleft{}Search for \ensuremath{\eta}' Bound Nuclei in the C12(\ensuremath{\gamma},p) Reaction with Simultaneous Detection of Decay Products\textquotedblright{},''
Phys. Rev. Lett. \textbf{126}, no.1, 019201 (2021)
[arXiv:2006.02912 [nucl-ex]].
 
\bibitem{n-PRiMESuper-FRS:2016vbn}
Y.~K.~Tanaka \textit{et al.} [n-PRiME/Super-FRS],
``Measurement of excitation spectra in the $^{12}$C$(p,d)$ reaction near the $\eta'$ emission threshold,''
Phys. Rev. Lett. \textbf{117}, no.20, 202501 (2016)
[arXiv:1611.02948 [nucl-ex]].

\bibitem{Bass:2013nya}
S.~D.~Bass and A.~W.~Thomas,
``QCD Symmetries in $\eta $- and $\eta '$-Mesic Nuclei,''
Acta Phys. Polon. B \textbf{45}, 627 (2014)
[arXiv:1311.7248 [hep-ph]].

\bibitem{Nagahiro:2011fi}
H.~Nagahiro, S.~Hirenzaki, E.~Oset and A.~Ramos,
``eta-prime nucleus optical potential and possible eta-prime bound states,''
Phys. Lett. B \textbf{709}, 87-92 (2012)
[arXiv:1111.5706 [hep-ph]].


\bibitem{Bernard:1987sx}
V.~Bernard and U.~G.~Meissner,
``Meson Properties at Finite Density From SU(3)-f Quark Dynamics,''
Phys. Rev. D \textbf{38}, 1551 (1988)

\bibitem{Nagahiro:2006dr}
H.~Nagahiro, M.~Takizawa and S.~Hirenzaki,
``eta- and eta-prime-mesic nuclei and U(A)(1) anomaly at finite density,''
Phys. Rev. C \textbf{74}, 045203 (2006)
[arXiv:nucl-th/0606052 [nucl-th]].

\bibitem{Costa:2002gk}
P.~Costa, M.~C.~Ruivo and Y.~L.~Kalinovsky,
``Pseudoscalar neutral mesons in hot and dense matter,''
Phys. Lett. B \textbf{560}, 171-177 (2003)
[arXiv:hep-ph/0211203 [hep-ph]].


\bibitem{Sakai:2013nba}
S.~Sakai and D.~Jido,
``In-medium $\eta^{\prime}$ mass and $\eta^{\prime}N$  interaction based on chiral effective theory,''
Phys. Rev. C \textbf{88}, no.6, 064906 (2013)
[arXiv:1309.4845 [nucl-th]].

\bibitem{Sakai:2022xao}
S.~Sakai and D.~Jido,
``Spectral function of the \ensuremath{\eta}' meson in nuclear medium based on phenomenological models,''
Phys. Rev. C \textbf{107}, no.2, 025207 (2023)
[arXiv:2212.05655 [nucl-th]].


\bibitem{Garcia-Recio:2002xog}
C.~Garcia-Recio, J.~Nieves, T.~Inoue and E.~Oset,
``eta bound states in nuclei,''
Phys. Lett. B \textbf{550}, 47-54 (2002)
[arXiv:nucl-th/0206024 [nucl-th]].

\bibitem{Inoue:2002xw}
T.~Inoue and E.~Oset,
``Eta in the nuclear medium within a chiral unitary approach,''
Nucl. Phys. A \textbf{710}, 354-370 (2002)
[arXiv:hep-ph/0205028 [hep-ph]].

\bibitem{Jido:2011pq}
D.~Jido, H.~Nagahiro and S.~Hirenzaki,
``Nuclear bound state of eta'(958) and partial restoration of chiral symmetry in the eta' mass,''
Phys. Rev. C \textbf{85}, 032201 (2012)
[arXiv:1109.0394 [nucl-th]].
 
 
\bibitem{Nagahiro:2004qz}
H.~Nagahiro and S.~Hirenzaki,
``Formation of eta-prime(958) - mesic nuclei and axial U(A)(1) anomaly at finite density,''
Phys. Rev. Lett. \textbf{94}, 232503 (2005)
[arXiv:hep-ph/0412072 [hep-ph]].


\bibitem{Tsushima:1998qw}
K.~Tsushima, D.~H.~Lu, A.~W.~Thomas and K.~Saito,
``Are eta and omega nuclear states bound?,''
Phys. Lett. B \textbf{443}, 26-32 (1998)
[arXiv:nucl-th/9806043 [nucl-th]].

\bibitem{Tsushima:1998qp}
K.~Tsushima,
``Study of omega - mesic, eta - mesic, eta-prime - mesic and D- - mesic nuclei,''
Nucl. Phys. A \textbf{670}, 198-201 (2000)
[arXiv:nucl-th/9901095 [nucl-th]].

\bibitem{Cobos-Martinez:2017vtr}
J.~J.~Cobos-Mart\'\i{}nez, K.~Tsushima, G.~Krein and A.~W.~Thomas,
``$\phi$ meson mass and decay width in nuclear matter and nuclei,''
Phys. Lett. B \textbf{771}, 113-118 (2017)
[arXiv:1703.05367 [nucl-th]].

\bibitem{Cobos-Martinez:2017woo}
J.~J.~Cobos-Mart\'\i{}nez, K.~Tsushima, G.~Krein and A.~W.~Thomas,
``$\Phi$-meson--nucleus bound states,''
Phys. Rev. C \textbf{96}, no.3, 035201 (2017)
[arXiv:1705.06653 [nucl-th]].

\bibitem{Cobos-Martinez:2020ynh}
J.~J.~Cobos-Mart\'\i{}nez, K.~Tsushima, G.~Krein and A.~W.~Thomas,
``$\eta_{c}$-nucleus bound states,''
Phys. Lett. B \textbf{811}, 135882 (2020)
[arXiv:2007.04476 [hep-ph]].

\bibitem{Zeminiani:2020aho}
G.~N.~Zeminiani, J.~J.~Cobos-Martinez and K.~Tsushima,
``$\varUpsilon $ and $\eta _b$ mass shifts in nuclear matter,''
Eur. Phys. J. A \textbf{57}, no.8, 259 (2021)
[arXiv:2012.11381 [hep-ph]].

\bibitem{Cobos-Martinez:2022fmt}
J.~J.~Cobos-Mart\'\i{}nez, G.~N.~Zeminiani and K.~Tsushima,
``\ensuremath{\Upsilon} and \ensuremath{\eta}b nuclear bound states,''
Phys. Rev. C \textbf{105}, no.2, 025204 (2022)
[arXiv:2201.05696 [nucl-th]].

\bibitem{Workman:2022ynf}
R.~L.~Workman \textit{et al.} [Particle Data Group],
``Review of Particle Physics,''
PTEP \textbf{2022}, 083C01 (2022)

\bibitem{Guichon:1987jp}
P.~A.~M.~Guichon,
``A Possible Quark Mechanism for the Saturation of Nuclear Matter,''
Phys. Lett. B \textbf{200}, 235-240 (1988)


\bibitem{Saito:2005rv}
K.~Saito, K.~Tsushima and A.~W.~Thomas,
``Nucleon and hadron structure changes in the nuclear medium and impact on observables,''
Prog. Part. Nucl. Phys. \textbf{58}, 1-167 (2007)
[arXiv:hep-ph/0506314 [hep-ph]].

\bibitem{Guichon:2018uew}
P.~A.~M.~Guichon, J.~R.~Stone and A.~W.~Thomas,
``Quark\textendash{}Meson-Coupling (QMC) model for finite nuclei, nuclear matter and beyond,''
Prog. Part. Nucl. Phys. \textbf{100}, 262-297 (2018)
[arXiv:1802.08368 [nucl-th]].

\bibitem{Tsushima:1997df}
K.~Tsushima, K.~Saito, A.~W.~Thomas and S.~V.~Wright,
``In-medium kaon and antikaon properties in the quark meson coupling model,''
Phys. Lett. B \textbf{429}, 239-246 (1998)
[erratum: Phys. Lett. B \textbf{436}, 453-453 (1998)]
[arXiv:nucl-th/9712044 [nucl-th]].

\bibitem{Guichon:1995ue}
P.~A.~M.~Guichon, K.~Saito, E.~N.~Rodionov and A.~W.~Thomas,
``The Role of nucleon structure in finite nuclei,''
Nucl. Phys. A \textbf{601}, 349 (1996).
[arXiv:nucl-th/9509034 [nucl-th]].

\bibitem{Tsushima:2020gun}
K.~Tsushima,
``Magnetic moments of the octet, decuplet, low-lying charm, and low-lying bottom baryons in a nuclear medium,''
PTEP \textbf{2022}, no.4, 043D02 (2022)
[arXiv:2008.03724 [hep-ph]].

\bibitem{Feldmann:1999uf}
T.~Feldmann,
``Quark structure of pseudoscalar mesons,''
Int. J. Mod. Phys. A \textbf{15}, 159-207 (2000)
[arXiv:hep-ph/9907491 [hep-ph]].


\bibitem{Saito:1996sf}
K.~Saito, K.~Tsushima and A.~W.~Thomas,
``Selfconsistent description of finite nuclei based on a relativistic quark model,''
Nucl. Phys. A \textbf{609}, 339-363 (1996)
[arXiv:nucl-th/9606020 [nucl-th]].

\bibitem{Saito:1997ae}
K.~Saito, K.~Tsushima and A.~W.~Thomas,
``Rho meson mass in light nuclei,''
Phys. Rev. C \textbf{56}, 566-569 (1997)
[arXiv:nucl-th/9703011 [nucl-th]].

\bibitem{Xie:2018aeg}
J.~J.~Xie, W.~H.~Liang and E.~Oset,
``$\eta$-$^{4}$He interaction from the $dd \rightarrow \eta^{4}$He reaction near threshold,''
Eur. Phys. J. A \textbf{55}, no.1, 6 (2019)
[arXiv:1805.12532 [nucl-th]].

 \end{thebibliography}
\end{document}